\documentclass[useAMS,usenatbib,usegraphicx]{mn2e}
\usepackage{lscape}
\usepackage{amssymb}
\renewcommand{\thefootnote}{\fnsymbol{footnote}}

\def\Oi{O{\sc i}}
\def\Cii{C{\sc ii}}
\def\Siii{Si{\sc ii}}
\def\Nii{N{\sc ii}}
\def\Oiii{O{\sc iii}}
\def\Niii{N{\sc iii}}

\title[Polarimetric study of Berkeley 59]{Broad-band Optical Polarimetric Studies toward the 
Galactic young star cluster Be 59}

\author[C. Eswaraiah et al.]
{C. Eswaraiah,$^{1}$\thanks{E-mail:eswar@aries.res.in}
A. K. Pandey,$^{1}$ G. Maheswar,$^{1}$ W. P. Chen,$^{2}$ D. K. Ojha$^{3}$  
\newauthor{and H. C. Chandola$^{4}$} \\ 
$^{1}$Aryabhatta Research Institute of Observational Sciences, Manora Peak, Nainital 263129, India\\
$^{2}$Institute of Astronomy, National Central University, Chung-Li 32054, Taiwan\\
$^{3}$Tata Institute of Fundamental Research, Mumbai 400 005, India\\
$^{4}$Department of Physics, Kumaun University, Nainital 263129, India}

\begin{document}

    \long\def\symbolfootnote[#1]#2{\begingroup%
    \def\thefootnote{\fnsymbol{footnote}}\footnote[#1]{#2}\endgroup} 

%

\date{Accepted$------$, Received$------$; in original form$------$}

\pagerange{\pageref{firstpage}--\pageref{lastpage}} \pubyear{2011}

\maketitle

\label{firstpage}

\begin{abstract}

We present multiwavelength optical linear polarimetric observations of 69 stars 
toward the young open cluster Be 59. The observations reveal the presence of three 
dust layers located at the distances of $\sim$300, $\sim$500 and $\sim$700 pc. 
The dust layers produce a total polarization $P_V \sim 5.5 $ per cent. The mean values of polarization 
and polarization angles due to the dust layers are found to increase systematically with distance. 
We show that polarimetry in combination with the $(U-B)-(B-V)$ colour-colour diagram yields a better identification 
of cluster members. The polarization measurements suggest that the polarization due the intra-cluster medium 
is $\sim$ 2.2 per cent.  An anomalous reddening law exists for the cluster region, indicating a relatively larger grain size 
than that in the diffuse ISM. The spatial variation of the polarization and $E(B-V)$ is found to 
increase with radial distance from the cluster center, whereas the $\theta_{V}$ and $\lambda_{max}$ are found to 
decrease with increasing radial distance from the cluster center. About 40 per cent of cluster members show 
the signatures of either intrinsic polarization 
or rotation in their polarization angles. There is an indication that the star light of the cluster 
members might have been depolarized because of non-uniform alignment of dust grains in the foreground dust layers 
and in the intra-cluster medium. 
\end{abstract}


\section{INTRODUCTION}

Interstellar grains are aspherical in nature and, given proper conditions, are 
aligned in space by the magnetic field
(Davis \& Greenstein 1951).
The effective extinction cross sections of the
dust particles are the greatest when the electric vector of
the incident light is parallel to the long axes of the
dust particles as projected on the plane of the sky, and
the least when parallel to the short axes. 
This differential extinction introduces a small degree of linear polarization
in the transmitted light. 

Studies of polarization due to the interstellar matter (ISM) are important 
as these provide information about the properties of 
the dust associated with the ISM and intra-cluster matter as well as help to 
trace the Galactic magnetic field. As the grains are thought to align due to the 
local magnetic field, the observed polarization vectors map the general geometry of 
the magnetic field. The observed maximum upper limit relation between 
the degree of the polarization and the colour 
excess $E(B-V)$ is found to be $P_{max}=9\times E(B-V)$ (Aannestad $\&$ Purcell 1973). 
The relation between $P_{max}$ and colour excess, and the variation of $P$ with wavelength 
are interpreted in terms of the grain properties and the efficiency of the grain alignment. 
Therefore, polarimetry is a useful technique to investigate the properties like maximum 
polarization $P_{\lambda_{max}}$, the wavelength $\lambda_{max}$ corresponding to 
$P_{\lambda_{max}}$ and the orientation of the magnetic field in various Galactic locations. 

Polarimetric studies of star-forming regions/young star clusters are specially important 
because physical parameters such as distance, age, membership and colour excess $E(B-V)$ of these 
regions are known accurately, which consequently helps in analyzing the polarimetric data 
in a meaningful way. Strong ultraviolet radiation from O/B type stars in these regions has strong impact 
on the surrounding medium. Dust grains can undergo destruction processes due to direct 
radiative pressure, grain-grain collisions, sputtering or shattering etc. As a result, 
it is likely that the mean size of the dust grains could be smaller than the mean value for the diffuse ISM. 
The stars associated with the star-forming regions can help to understand the nature of dust as well 
as the magnetic field of the intra-cluster medium. 

Young star clusters (age $\textless$ 10 Myr), still embedded in the parent 
molecular clouds, are unique laboratories to understand the dust properties as well as the nature 
of interaction between young star(s) and the surrounding medium. 
Berkeley 59 ($\alpha_{2000}$ = 00h 02m 13s, $\delta_{2000}$= 
+67$\degr$ 25$^\prime$ 11$^\prime$$^\prime$; $l$ = 118$\degr$.22, $b$ = 5$\degr$.00) is such a young star 
cluster associated with a heavily obscured gas-dust complex of the Cepheus OB4 association. 
The cluster Be 59, located at the center of the 
Sharpless region S171 contains nine O7$-$B3 stars (cf. Pandey et al. 2008; hereafter P08) 
at a distance of  $1.00\pm0.05$ kpc and has $E(B-V)$ $\simeq$ $1.4-1.8$ mag. 
The extent of the cluster was found to be $2.9$ pc (P08).

As a part of an ongoing project to understand the dust characteristics in star-forming regions 
and to map the structure of magnetic field at diverse environments of the Milky Way Galaxy, 
we have carried out broad-band optical polarimetric observations around the cluster Be 59.
In Section 2 we present the observations and data reduction. Results are presented in Section 3. 
The dust properties and the spatial variation of $E(B-V)$, $P_V$, $\theta_V$ and $\lambda_{max}$ 
are in Sections 4 and 5 respectively. Finally, we conclude our 
results in Section 6.

\section{OBSERVATIONS AND  DATA REDUCTION}
Polarimetric observations were carried out
on seven nights (2009 November 23, 24, 25
and 2009 December 24, 26, 27 and 28), using the ARIES Imaging Polarimeter
(AIMPOL; Rautela, Joshi \& Pandey, 2004)
mounted at the Cassegrain
focus of the 104-cm Sampurnanand telescope of the Aryabhatta
Research Institute of observational sciencES (ARIES), Nainital, India.
The observations were carried out in the $B$, $V$, $R_{c}$ and $I_{c}$
($\lambda_{B_{eff}}$=0.440$\mu$m, $\lambda_{V_{eff}}$=0.530$\mu$m,
$\lambda_{Rc_{eff}}$=0.670$\mu$m and $\lambda_{I_{eff}}$=0.800$\mu$m)
photometric bands using fraction (370$\times$370 pixel$^2$)
of the TK 1024 $\times$ 1024 pixel$^2$ CCD camera.
The AIMPOL consists of a half-wave plate modulator and a Wollaston prism beam-splitter.
The Wollaston prism analyzer is placed at
the backend of the telescope beam path in order to produce
ordinary and extraordinary beams in slightly different directions separated by 28 pixels 
along the North-South direction on the sky plane.
A focal reducer (85 mm, f/1.8) is placed between the Wollaston prism and the CCD camera.
Each pixel of the CCD corresponds to 1.73 arcsec
and the field of view is $\sim$~8 arcmin
diameter on the sky. The FWHM of the stellar image varied from 2 to 3 pixels.
The read-out noise and gain of the CCD are
7.0 $e^{-1}$  and 11.98 $e^{-1}$/ADU respectively.
Since AIMPOL does not have a grid, we manually checked for any overlap of ordinary 
and extraordinary images of the sources.
Fluxes of ordinary ($I_{o}$) and extra-ordinary ($I_{e}$) beams for all the observed sources
with good signal-to-noise ratio were extracted by standard aperture photometry
after bias subtraction using the {\small IRAF}\symbolfootnote[1]{{\small IRAF} is distributed by National Optical
Astronomical Observatories, USA.} package.
The ratio $R(\alpha)$ is computed using the following relation 
\begin{equation}\label{R_alpha}
 R(\alpha) = \frac{\frac{{I_{e}}(\alpha)}{{I_{o}}(\alpha)}-1} {\frac{I_{e}(\alpha)} 
{I_{o}(\alpha)}+1} =  P cos(2\theta - 4\alpha),
\end{equation}
where $P$ is the fraction of the total light in the linearly polarized condition and $\theta$
is the position angle of the plane of polarization.
Here $\alpha$ is the position of the
fast axis of the half wave plate (HWP) at
0$\degr$, 22.5$\degr$, 45$\degr$ and 67.5$\degr$ corresponding to the four
normalized Stokes parameters respectively, $q$ [R(0$\degr$)], $u$ [R(22.5$\degr$)],
$q_{1}$ [R(45$\degr$)] and $u_{1}$ [R(67.5$\degr$)].
The detailed procedures used to estimate the polarization and position
angles for the program stars are given by Eswaraiah et al. (2011) (hereafter E11) and references therein. 

The instrumental polarization of AIMPOL on the 104-cm Sampurnanand
Telescope has been monitored since 2004 
on various observing nights 
and found to be less than 0.1 per cent in different bands 
(E11 and references therein). 
All the measurements were corrected for both the null polarization ($\sim$ 0.1 per cent)
which is independent of the pass-bands and the zero-point polarization angle
by observing several unpolarized and polarized standard stars
from Schmidt, Elston \& Lupie (1992) (hereafter S92).
The results for the standard stars are given in Table \ref{stand_results}.
The first column lists the star name with the
date of observation or reference in parenthesis. The next consecutive columns are the polarization
in percent [$P$ (per cent)] and polarization angle in degree [$\theta (\degr)$]
measured in BV$(RI)_{C}$ pass-bands. Entries with S92 in  parenthesis are taken from S92. 
The present results for the polarized standard stars
are in good agreement, within the observational errors, with those given by S92.

\section{RESULTS}
Table \ref{BVRI_pol_data} lists the polarization measurements for 37 stars using $BV(RI)_{C}$
bands, whereas Table \ref{VRI_pol_data} lists the results for 32 stars using the $V(RI)_{C}$ bands.
The star identification numbers (column 1) are taken from P08.
The right ascension, declination and
photometric visual magnitudes, also from P08, are listed in
the second, third and fourth columns of Tables
\ref{BVRI_pol_data} and \ref{VRI_pol_data} respectively.
The next consecutive columns correspond to the polarization, polarization angle
and their associated errors in B, V, $(R,I)_{c}$ bands and V, $(R,I)_{c}$ bands respectively.
The given polarization angles are in the equatorial coordinate system
measured from the North increasing towards the East.
Tables \ref{BVRI_pol_data} and \ref{VRI_pol_data} 
reveal that the maximum linear polarization towards the cluster region is $\sim$ 8 per cent.
Such a high amount of polarization is not often
found towards the star clusters, with only a few exceptions, $e.g.,$ Tr 27 (Feinstein et al. 2000),
and M17 (Schultz et al. 1981).

\begin{figure*}
\vskip.1cm
\resizebox{17cm}{16cm}{\includegraphics{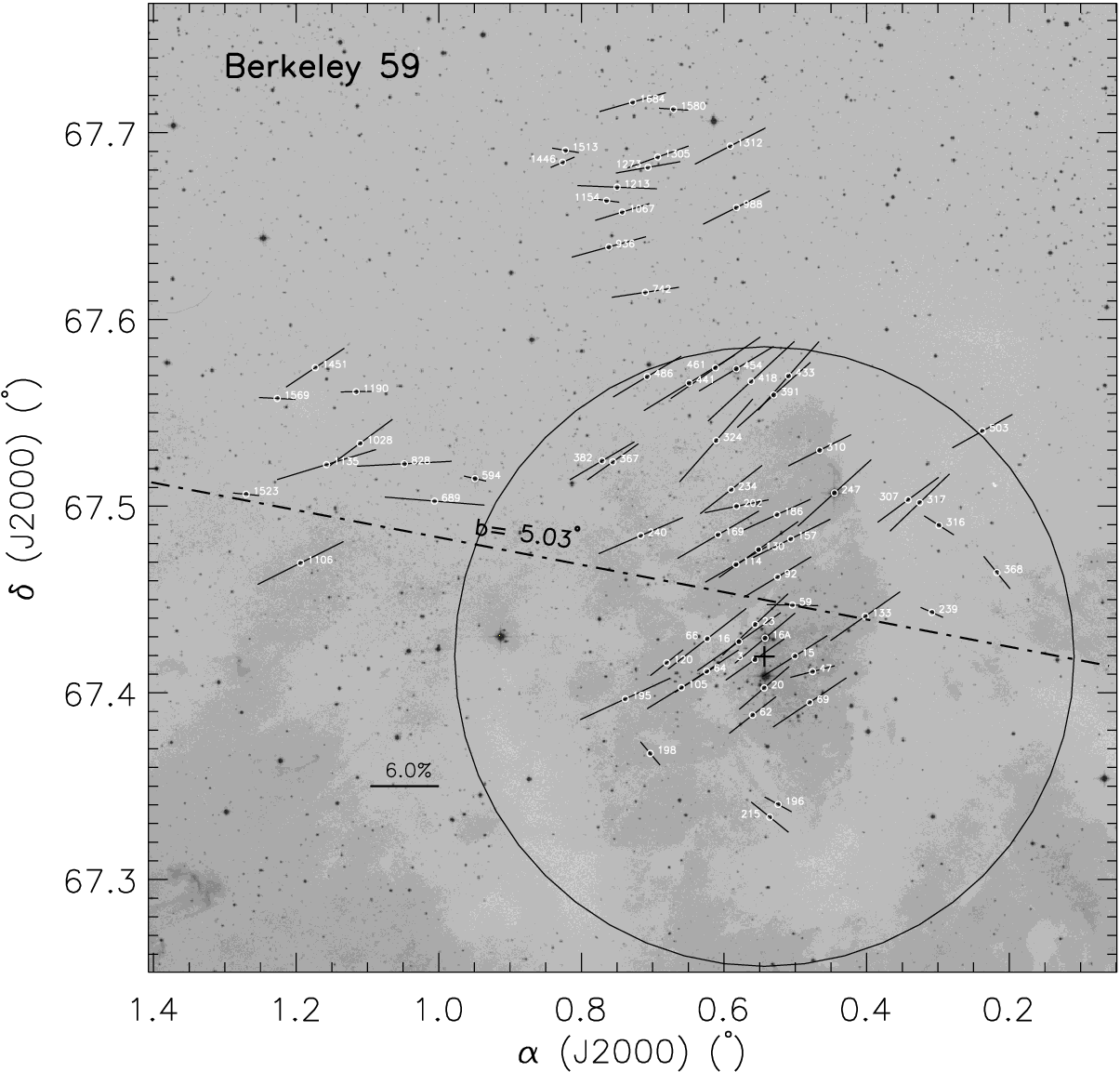}}
\caption{The stellar polarization vectors are superimposed on a 21$^\prime \times$ 19$^\prime$
R-band DSS II image of the
field containing Be 59.
The length of the polarization vector is proportional to $P_{V}$.
A vector with a polarization of 6 per cent is drawn for
reference. The dash-dotted line is the Galactic plane at b=5.03$\degr$.
The plus symbols represents the cluster center
RA=00$^{h}$02$^{m}$10$^{s}$.4 and
Dec=67$\degr$25$^\prime$10$^\prime$$^\prime$.
The cluster radius is shown with a big circle of $\sim$10 arcmin (P08).
The stars observed are identified using the star identification numbers from P08.
North is at the top and east is to the left.}
\label{starids_Be59}
\end{figure*}

In Fig. \ref{starids_Be59} all the observed 69 stars are marked with white
circles on the DSS II $R$-band image.
The sky projection of $V$-band polarization vectors
are overlaid. The length of each polarization vector is proportional to the
degree of polarization. 
The dash-dotted line represents the orientation of the projection of the Galactic plane (GP) at $b$=5.03$\degr$, 
which corresponds to the position angle of $86\degr$.

\begin{landscape}
\begin{table}
\centering
\small
\caption{Observed polarized standard stars from S92}\label{stand_results}
\begin{tabular}{cccccccccc}\hline \hline
star name & $P_{B}\pm\epsilon$ & $\theta_{B}\pm\epsilon$ & $P_{V}\pm\epsilon$ & $\theta_{V}\pm\epsilon$ & $P_{Rc}\pm\epsilon$ & $\theta_{Rc}\pm\epsilon$ & $P_{Ic}\pm\epsilon$ & $\theta_{Ic}\pm\epsilon$ \\
(date of observations) &&&&&&&&   \\
(reference) & (per cent)  &  ($\degr$)  & (per cent) &  ($\degr$) & (per cent) & ($\degr$)  & (per cent)  &  ($\degr$)  \\
 (1)        &  (2)    &  (3)         &   (4)  &  (5)        &   (6)   &  (7)        &  (8)    &   (9)    \\
\hline
\hline
HD 19820 (23, Nov 09) & 4.49 $\pm$ 0.11 & 114.9 $\pm$ 0.7 & 4.89  $\pm$ 0.09 & 114.2 $\pm$ 0.5 & 4.49 $\pm$ 0.09  & 115.5 $\pm$ 0.6 & 4.06 $\pm$ 0.16  & 115.3 $\pm$ 1.0  \\ 
HD 19820 (24, Nov 09) & 4.61 $\pm$ 0.11 & 114.7 $\pm$ 0.7 & 4.79  $\pm$ 0.08 & 115.1 $\pm$ 0.5 & 4.51 $\pm$ 0.07  & 114.6 $\pm$ 0.4 & 3.97 $\pm$ 0.09  & 115.7 $\pm$ 0.6  \\ 
HD 19820 (25, Nov 09) & 4.72 $\pm$ 0.11 & 115.6 $\pm$ 0.7 & 4.94  $\pm$ 0.08 & 114.9 $\pm$ 0.4 & 4.61 $\pm$ 0.07  & 114.4 $\pm$ 0.4 & 3.86 $\pm$ 0.10  & 113.4 $\pm$ 0.7  \\
HD 19820 (23, Dec 09) & 4.70 $\pm$ 0.10 & 115.0 $\pm$ 0.6 &       ---       &       ---       & 4.61 $\pm$ 0.07  & 114.3 $\pm$ 0.4 & 4.06 $\pm$ 0.08  & 114.0 $\pm$ 0.5  \\ 
HD 19820 (24, Dec 09) & 4.68 $\pm$ 0.11 & 116.4 $\pm$ 0.7 & 4.90  $\pm$ 0.09 & 114.6 $\pm$ 0.5 & 4.62 $\pm$ 0.07  & 114.8 $\pm$ 0.4 & 4.06 $\pm$ 0.08  & 114.3 $\pm$ 0.5  \\
HD 19820 (27, Dec 09) & 4.57 $\pm$ 0.11 & 116.4 $\pm$ 0.6 & 4.69  $\pm$ 0.09 & 115.0 $\pm$ 0.5 & 4.63 $\pm$ 0.07  & 114.6 $\pm$ 0.4 & 4.01 $\pm$ 0.08  & 114.3 $\pm$ 0.5  \\
HD 19820 (28, Dec 09) & 4.72 $\pm$ 0.09 & 115.4 $\pm$ 0.5 & 4.80  $\pm$ 0.08 & 114.8 $\pm$ 0.5 & 4.51 $\pm$ 0.07  & 114.2 $\pm$ 0.4 & 3.92 $\pm$ 0.08  & 115.2 $\pm$ 0.5  \\
HD 19820 (S92)        & 4.70 $\pm$ 0.04 & 115.7 $\pm$ 0.2 & 4.79 $\pm$ 0.03 & 114.9 $\pm$  0.2 & 4.53 $\pm$ 0.03  & 114.5 $\pm$ 0.2 & 4.08 $\pm$0.02  & 114.5 $\pm$ 0.2   \\
\hline
HD 25443  (23, Nov 09) & 5.19 $\pm$ 0.09 & 134.6 $\pm$ 0.5 & 5.04 $\pm$ 0.07 & 136.0 $\pm$ 0.4 & 5.01 $\pm$ 0.06 & 134.5 $\pm$ 0.4 & 4.19 $\pm$ 0.09 & 134.8 $\pm$ 0.6 \\
HD 25443  (24, Nov 09) & 5.21 $\pm$ 0.09 & 134.1 $\pm$ 0.5 & 5.11 $\pm$ 0.07 & 136.2 $\pm$ 0.4 & 5.13 $\pm$ 0.06 & 133.8 $\pm$ 0.4 & 4.29 $\pm$ 0.09 & 134.4 $\pm$ 0.6 \\
HD 25443  (25, Nov 09) & 5.11 $\pm$ 0.09 & 134.5 $\pm$ 0.5 & 5.25 $\pm$ 0.09 & 134.3 $\pm$ 0.5 & 5.10 $\pm$ 0.08 & 133.4 $\pm$ 0.4 & 4.37 $\pm$ 0.10 & 132.6 $\pm$ 0.7 \\
HD 25443  (23, Dec 09) & 5.12 $\pm$ 0.08 & 135.7 $\pm$ 0.5 & 5.12 $\pm$ 0.08 & 135.5 $\pm$ 0.4 & 4.88 $\pm$ 0.08 & 134.5 $\pm$ 0.5 & 4.23 $\pm$ 0.08 & 134.9 $\pm$ 0.5 \\
HD 25443  (24, Dec 09) & 5.12 $\pm$ 0.09 & 134.0 $\pm$ 0.5 & 5.27 $\pm$ 0.09 & 134.8 $\pm$ 0.5 & 5.00 $\pm$ 0.08 & 134.5 $\pm$ 0.4 & 4.19 $\pm$ 0.07 & 136.0 $\pm$ 0.5  \\
HD 25443  (27, Dec 09) & 5.05 $\pm$ 0.08 & 134.8 $\pm$ 0.5 & 5.17 $\pm$ 0.07 & 135.9 $\pm$ 0.4 & 5.04 $\pm$ 0.07 & 134.0 $\pm$ 0.4 & 4.13 $\pm$ 0.07 & 134.5 $\pm$ 0.5  \\
HD 25443 (S92)         & 5.23 $\pm$ 0.09 & 134.3 $\pm$ 0.5 & 5.13 $\pm$ 0.06 & 134.2 $\pm$ 0.3 & 4.73 $\pm$ 0.05 & 133.6 $\pm$ 0.3 & 4.25 $\pm$ 0.04 & 134.2 $\pm$ 0.3 \\
\hline
BD+64 106 (23, Nov 09) & 5.49 $\pm$ 0.17 & 98.0 $\pm$ 0.9  & 6.09 $\pm$ 0.13 & 99.0 $\pm$ 0.6 & 5.41 $\pm$ 0.11 & 96.1 $\pm$ 0.6 & 4.50 $\pm$ 0.14 & 96.6 $\pm$ 0.9  \\
BD+64 106(23, Dec 09)  & 5.37 $\pm$ 0.15 & 97.9 $\pm$ 0.8  & 5.57 $\pm$ 0.11 & 96.8 $\pm$ 0.6 & 5.63 $\pm$ 0.09 & 97.6 $\pm$ 0.5 & 4.63 $\pm$ 0.10 & 97.5 $\pm$ 0.6 \\
BD+64 106 (S92)        & 5.51 $\pm$ 0.09 & 97.2 $\pm$ 0.5  & 5.69 $\pm$ 0.04 & 96.6 $\pm$ 0.2 & 5.15 $\pm$ 0.10 & 96.7 $\pm$ 0.5 & 4.70 $\pm$ 0.05 & 96.9 $\pm$ 0.3 \\
\hline
HD 204827 (25, Nov 09) & 5.67 $\pm$ 0.10 & 58.3 $\pm$ 0.5 & 5.53 $\pm$ 0.08 & 58.8 $\pm$ 0.4 & 5.02 $\pm$ 0.08 & 59.2 $\pm$ 0.4 & 4.14 $\pm$ 0.10 & 59.9 $\pm$ 0.7 \\
HD 204827 (S92)        & 5.65 $\pm$ 0.02 & 58.2 $\pm$ 0.1 & 5.32 $\pm$ 0.01 & 58.7 $\pm$ 0.1 & 4.89 $\pm$ 0.03 & 59.1 $\pm$ 0.2 & 4.19 $\pm$ 0.03 & 59.9 $\pm$ 0.2 \\
\hline
BD+59 389 (23, Dec 09) & 6.35 $\pm$ 0.13  & 97.8 $\pm$ 0.6  & 6.73 $\pm$ 0.09  & 97.8 $\pm$ 0.4  & 6.48 $\pm$ 0.08  & 97.6 $\pm$ 0.3  & 5.66 $\pm$ 0.06  & 97.9 $\pm$ 0.3  \\
BD+59 389 (28, Dec 09) & 6.43 $\pm$ 0.13  & 97.9 $\pm$ 0.6  & 6.82 $\pm$ 0.09  & 97.5 $\pm$ 0.4  & 6.47 $\pm$ 0.08  & 97.7 $\pm$ 0.4  & 5.61 $\pm$ 0.07  & 97.9 $\pm$ 0.4 \\
BD+59 389 (S92)        & 6.34 $\pm$ 0.04  & 98.1 $\pm$ 0.2  & 6.70 $\pm$ 0.01  & 98.1 $\pm$ 0.1  & 6.43 $\pm$ 0.02  & 98.1 $\pm$ 0.1  & 5.80 $\pm$ 0.02  & 98.3 $\pm$ 0.1 \\
\hline
HD 236633 (28, Dec 09) & 6.06 $\pm$ 0.11  & 90.4 $\pm$ 0.5  & 5.65 $\pm$ 0.09  & 91.3 $\pm$ 0.4  & 5.34 $\pm$ 0.09 & 91.0 $\pm$ 0.5 & 4.69 $\pm$ 0.09 & 90.5 $\pm$ 0.6   \\
HD 236633 (S92)        & 5.53 $\pm$ 0.04  & 92.5 $\pm$ 0.2  & 5.49 $\pm$ 0.02  & 93.8 $\pm$ 0.1  & 5.38 $\pm$ 0.03 & 93.0 $\pm$ 0.2 & 4.80 $\pm$ 0.04 & 93.1 $\pm$ 0.2 \\
\hline
\hline
\end{tabular}
\end{table}
\end{landscape}

\begin{landscape}
\begin{table}
\centering
\small
\caption{Observed BV$(RI)_{c}$ polarization and polarization angles for 37 stars towards Be 59}
\label{BVRI_pol_data}
\begin{tabular}{cccccccccccc}\hline \hline
Star ID $^\star$ &  R.A  ($\degr$)   & DEC ($\degr$)   &  V (mag)$^\star$  & $P_{B}\pm \epsilon$ & $\theta_{B}\pm\epsilon$ &  $P_{V}\pm\epsilon$& $\theta_{V}\pm\epsilon$ & $P_{Rc}\pm\epsilon$
& $\theta_{Rc}\pm\epsilon$ & $P_{Ic}\pm\epsilon $& $\theta_{Ic}\pm\epsilon$  \\
 &   (2000J)  & (2000J) &  & (per cent) & ($\degr$) & (per cent) & ($\degr$) & (per cent) & ($\degr$) & (per cent) & ($\degr$)   \\
 (1)   &  (2)     &   (3)    &   (4)  &  (5)   &   (6)   &  (7)      &  (8)        &   (9)  &  (10)  &  (11) &  (12) \\
\hline
    
  3   &  0.55659104  &    67.417798  &  11.30  &   4.80 $\pm$  0.20  &   102.3 $\pm$   1.2  &   5.30 $\pm$  0.13  &   105.1  $\pm$  0.7  &   5.00 $\pm$  0.10  &   104.0  $\pm$  0.5  &   4.57 $\pm$  0.09  &   106.7 $\pm$   0.6 \\ 
  15  &  0.50060639  &    67.419664  &  12.78  &   5.04 $\pm$  0.41  &   103.4 $\pm$   2.3  &   5.85 $\pm$  0.26  &   103.6  $\pm$  1.2  &   6.20 $\pm$  0.19  &   103.4  $\pm$  0.9  &   5.48 $\pm$  0.18  &   104.4 $\pm$   0.9 \\ 
  16  &  0.57895372  &    67.427389  &  11.81  &   6.87 $\pm$  0.31  &   106.8 $\pm$   1.3  &   7.55 $\pm$  0.16  &   104.9  $\pm$  0.6  &   7.04 $\pm$  0.09  &   106.0  $\pm$  0.4  &   6.68 $\pm$  0.07  &   107.7 $\pm$   0.3 \\ 
 16A  &  0.54241389  &    67.429288  &    -    &   5.35 $\pm$  0.09  &   106.3 $\pm$   0.5  &   5.53 $\pm$  0.06  &   107.7  $\pm$  0.3  &   5.45 $\pm$  0.04  &   106.4  $\pm$  0.2  &   4.64 $\pm$  0.04  &   106.5 $\pm$   0.3 \\ 
  20  &  0.54363477  &    67.402592  &  13.40  &   3.78 $\pm$  0.47  &   108.1 $\pm$   3.5  &   4.60 $\pm$  0.31  &   108.2  $\pm$  1.9  &   4.21 $\pm$  0.23  &   106.3  $\pm$  1.5  &   3.76 $\pm$  0.21  &   109.0 $\pm$   1.6 \\ 
  23  &  0.55624163  &    67.436586  &  13.84  &   6.85 $\pm$  0.75  &   101.9 $\pm$   3.1  &   6.10 $\pm$  0.42  &   109.7  $\pm$  1.9  &   5.97 $\pm$  0.27  &   112.9  $\pm$  1.3  &   5.57 $\pm$  0.23  &   110.9 $\pm$   1.2 \\ 
  66  &  0.62357699  &    67.428952  &  12.95  &   6.81 $\pm$  0.50  &   107.5 $\pm$   2.1  &   7.14 $\pm$  0.28  &   106.5  $\pm$  1.1  &   6.99 $\pm$  0.19  &   107.7  $\pm$  0.8  &   6.34 $\pm$  0.16  &   108.1 $\pm$   0.7 \\ 
  69  &  0.47981658  &    67.394837  &  14.06  &   6.16 $\pm$  0.74  &   111.3 $\pm$   3.3  &   6.62 $\pm$  0.47  &   104.4  $\pm$  2.0  &   6.74 $\pm$  0.35  &   107.5  $\pm$  1.5  &   5.39 $\pm$  0.33  &   103.3 $\pm$   1.7 \\ 
 114  &  0.58320121  &    67.468772  &  12.43  &   5.37 $\pm$  0.28  &    97.8 $\pm$   1.5  &   5.50 $\pm$  0.17  &   102.5  $\pm$  0.9  &   5.39 $\pm$  0.11  &    96.7  $\pm$  0.6  &   4.66 $\pm$  0.13  &    98.3 $\pm$   0.8 \\ 
 130  &  0.55160126  &    67.476766  &  12.61  &   5.84 $\pm$  0.34  &    99.0 $\pm$   1.6  &   7.17 $\pm$  0.19  &   105.1  $\pm$  0.7  &   6.80 $\pm$  0.11  &    98.4  $\pm$  0.5  &   6.14 $\pm$  0.11  &    98.2 $\pm$   0.5 \\ 
 157  &  0.50636486  &    67.482452  &  14.51  &   6.89 $\pm$  0.96  &    94.8 $\pm$   3.9  &   7.06 $\pm$  0.57  &   100.8  $\pm$  2.3  &   6.62 $\pm$  0.40  &    99.0  $\pm$  1.7  &   5.74 $\pm$  0.35  &    94.6 $\pm$   1.7 \\ 
 169  &  0.60818307  &    67.484653  &  13.43  &   6.86 $\pm$  0.48  &    95.0 $\pm$   2.0  &   7.29 $\pm$  0.28  &   102.6  $\pm$  1.1  &   7.16 $\pm$  0.17  &    99.2  $\pm$  0.7  &   6.34 $\pm$  0.19  &   100.5 $\pm$   0.8 \\ 
 186  &  0.52564189  &    67.495438  &  13.72  &   5.44 $\pm$  0.67  &    99.7 $\pm$   3.5  &   5.95 $\pm$  0.36  &    99.9  $\pm$  1.7  &   5.70 $\pm$  0.20  &   100.3  $\pm$  1.0  &   5.00 $\pm$  0.19  &   101.0 $\pm$   1.1 \\ 
 234  &  0.58994890  &    67.508870  &  14.22  &   4.81 $\pm$  0.65  &    96.4 $\pm$   3.8  &   5.66 $\pm$  0.39  &   106.6  $\pm$  1.9  &   5.16 $\pm$  0.26  &   101.3  $\pm$  1.4  &   4.57 $\pm$  0.31  &    99.0 $\pm$   1.9 \\ 
 239  &  0.30857984  &    67.443113  &  12.94  &   2.46 $\pm$  0.43  &    81.1 $\pm$   4.8  &   1.98 $\pm$  0.28  &    80.4  $\pm$  3.5  &   2.24 $\pm$  0.18  &    82.7  $\pm$  2.2  &   1.78 $\pm$  0.21  &    83.5 $\pm$   3.2 \\ 
 247  &  0.44508728  &    67.507038  &  10.09  &   6.14 $\pm$  0.12  &   106.7 $\pm$   0.6  &   6.80 $\pm$  0.07  &   109.1  $\pm$  0.3  &   6.48 $\pm$  0.05  &   110.4  $\pm$  0.2  &   5.87 $\pm$  0.05  &   112.0 $\pm$   0.2 \\ 
 307  &  0.34187817  &    67.503367  &  13.97  &   4.73 $\pm$  0.97  &   103.0 $\pm$   5.8  &   5.55 $\pm$  0.45  &   105.7  $\pm$  2.3  &   5.43 $\pm$  0.27  &   107.1  $\pm$  1.4  &   5.05 $\pm$  0.22  &   108.3 $\pm$   1.2 \\ 
 310  &  0.46585155  &    67.529885  &  12.48  &   6.39 $\pm$  0.34  &   101.4 $\pm$   1.5  &   5.59 $\pm$  0.20  &   100.8  $\pm$  1.0  &   5.77 $\pm$  0.12  &   101.9  $\pm$  0.6  &   4.89 $\pm$  0.12  &   102.8 $\pm$   0.7 \\ 
 316  &  0.29854431  &    67.489695  &  12.47  &   3.14 $\pm$  0.36  &    74.9 $\pm$   3.2  &   2.67 $\pm$  0.27  &    76.3  $\pm$  2.5  &   2.92 $\pm$  0.19  &    75.3  $\pm$  1.8  &   2.17 $\pm$  0.24  &    66.9 $\pm$   3.0 \\ 
 324  &  0.61096275  &    67.535161  &  14.13  &   7.21 $\pm$  0.65  &   111.1 $\pm$   2.6  &   6.98 $\pm$  0.38  &   113.4  $\pm$  1.5  &   6.63 $\pm$  0.24  &   109.7  $\pm$  1.0  &   6.00 $\pm$  0.27  &   110.1 $\pm$   1.3 \\ 
 367  &  0.75596949  &    67.523695  &  14.88  &   3.74 $\pm$  1.00  &    88.1 $\pm$   7.5  &   4.64 $\pm$  0.54  &   105.1  $\pm$  3.3  &   5.95 $\pm$  0.33  &    99.6  $\pm$  1.6  &   5.89 $\pm$  0.36  &   101.2 $\pm$   1.7 \\ 
 368  &  0.21731676  &    67.464417  &  13.93  &   3.10 $\pm$  0.67  &    73.8 $\pm$   6.0  &   2.54 $\pm$  0.43  &    64.8  $\pm$  4.4  &   2.72 $\pm$  0.29  &    76.7  $\pm$  3.0  &   2.55 $\pm$  0.33  &    75.7 $\pm$   3.6 \\ 
 382  &  0.77096790  &    67.524242  &  14.09  &   5.53 $\pm$  0.57  &   100.7 $\pm$   2.9  &   5.80 $\pm$  0.37  &   102.7  $\pm$  1.8  &   5.88 $\pm$  0.25  &   100.9  $\pm$  1.2  &   5.23 $\pm$  0.31  &   102.4 $\pm$   1.6 \\ 
 391  &  0.53020318  &    67.559531  &  13.15  &   6.61 $\pm$  0.41  &   106.4 $\pm$   1.8  &   6.76 $\pm$  0.24  &   108.8  $\pm$  1.0  &   6.86 $\pm$  0.15  &   104.1  $\pm$  0.6  &   6.13 $\pm$  0.17  &   103.6 $\pm$   0.8 \\ 
 418  &  0.56191788  &    67.566861  &  13.94  &   7.03 $\pm$  0.59  &   104.4 $\pm$   2.4  &   8.03 $\pm$  0.35  &   109.4  $\pm$  1.2  &   7.58 $\pm$  0.24  &   103.6  $\pm$  0.9  &   6.75 $\pm$  0.29  &   103.1 $\pm$   1.2 \\ 
 441  &  0.64887476  &    67.565940  &  12.14  &   7.48 $\pm$  0.25  &    98.7 $\pm$   0.9  &   8.15 $\pm$  0.15  &   103.2  $\pm$  0.5  &   7.91 $\pm$  0.10  &    99.1  $\pm$  0.4  &   7.16 $\pm$  0.12  &    98.5 $\pm$   0.5 \\ 
 454  &  0.58273633  &    67.573569  &  11.07  &   6.81 $\pm$  0.14  &    99.1 $\pm$   0.6  &   6.71 $\pm$  0.09  &   102.9  $\pm$  0.4  &   6.67 $\pm$  0.06  &    98.1  $\pm$  0.3  &   5.78 $\pm$  0.08  &    99.2 $\pm$   0.4 \\ 
 461  &  0.61191189  &    67.574077  &  13.72  &   7.45 $\pm$  0.52  &    96.6 $\pm$   2.0  &   8.10 $\pm$  0.32  &   104.7  $\pm$  1.1  &   7.45 $\pm$  0.21  &   100.6  $\pm$  0.8  &   6.42 $\pm$  0.25  &   101.4 $\pm$   1.1 \\ 
 486  &  0.70776450  &    67.569370  &  13.51  &   5.86 $\pm$  0.71  &    97.1 $\pm$   3.4  &   6.13 $\pm$  0.28  &   102.9  $\pm$  1.3  &   5.78 $\pm$  0.14  &    98.4  $\pm$  0.7  &   4.99 $\pm$  0.13  &   100.6 $\pm$   0.7 \\ 
 936  &  0.76152588  &    67.638797  &  12.63  &   6.39 $\pm$  0.36  &    96.2 $\pm$   1.6  &   6.53 $\pm$  0.18  &    96.0  $\pm$  0.8  &   6.23 $\pm$  0.12  &    95.8  $\pm$  0.5  &   5.41 $\pm$  0.13  &    94.6 $\pm$   0.6 \\ 
 988  &  0.58255340  &    67.659929  &  13.13  &   5.82 $\pm$  0.46  &   102.7 $\pm$   2.2  &   5.93 $\pm$  0.23  &   100.9  $\pm$  1.1  &   5.48 $\pm$  0.17  &    99.2  $\pm$  0.9  &   4.62 $\pm$  0.19  &   102.5 $\pm$   1.2 \\ 
1067  &  0.74265875  &    67.657519  &  13.43  &   3.94 $\pm$  0.50  &    99.3 $\pm$   3.5  &   4.76 $\pm$  0.26  &    96.8  $\pm$  1.5  &   4.36 $\pm$  0.18  &    97.2  $\pm$  1.1  &   3.79 $\pm$  0.20  &    98.1 $\pm$   1.4 \\ 
1154  &  0.76472099  &    67.663676  &  13.54  &   1.94 $\pm$  0.49  &    87.7 $\pm$   6.9  &   2.20 $\pm$  0.27  &    87.1  $\pm$  3.4  &   2.02 $\pm$  0.20  &    83.8  $\pm$  2.7  &   1.53 $\pm$  0.24  &    85.2 $\pm$   4.2 \\ 
1213  &  0.74994583  &    67.670815  &  13.36  &   6.57 $\pm$  0.55  &    88.9 $\pm$   2.4  &   6.97 $\pm$  0.26  &    89.1  $\pm$  1.1  &   6.53 $\pm$  0.17  &    89.2  $\pm$  0.7  &   5.40 $\pm$  0.17  &    90.2 $\pm$   0.9 \\ 
1446  &  0.82639705  &    67.684169  &  14.26  &   2.32 $\pm$  0.71  &   106.3 $\pm$   8.4  &   2.11 $\pm$  0.39  &    99.5  $\pm$  5.0  &   1.90 $\pm$  0.28  &    95.2  $\pm$  4.0  &   1.76 $\pm$  0.34  &    89.3 $\pm$   5.1 \\ 
1580  &  0.67072198  &    67.712589  &  12.93  &   2.01 $\pm$  0.38  &    90.7 $\pm$   5.2  &   2.55 $\pm$  0.21  &    88.5  $\pm$  2.2  &   2.34 $\pm$  0.15  &    92.4  $\pm$  1.8  &   2.17 $\pm$  0.18  &    92.7 $\pm$   2.2 \\ 
1684  &  0.72806296  &    67.716366  &  13.27  &   6.45 $\pm$  0.68  &    92.5 $\pm$   3.0  &   5.84 $\pm$  0.25  &    96.2  $\pm$  1.2  &   5.52 $\pm$  0.13  &    96.7  $\pm$  0.7  &   4.57 $\pm$  0.12  &    98.0 $\pm$   0.7 \\ 
\hline \hline                                                                                                    
\end{tabular}\\
\hspace{-10cm}
$^\star$ Star IDs and V magnitudes, barring 16A, have been taken from P08
\end{table}
\end{landscape}

\begin{landscape}
\begin{table}
\centering
\small
\caption{Observed V, $(RI)_{c}$ polarization and polarization angles for 32 stars towards Be 59}
\label{VRI_pol_data}
\begin{tabular}{cccccccccc}\hline \hline
Star ID $^\star$
&  R.A  ($\degr$)   & DEC ($\degr$)   &  V (mag)$^\star$ &  $P_{V}\pm\epsilon$& $\theta_{V}\pm\epsilon$ & $P_{Rc}\pm\epsilon$
& $\theta_{Rc}\pm\epsilon$ & $P_{Ic}\pm\epsilon $& $\theta_{Ic}\pm\epsilon$  \\
  &   (2000J)  & (2000J) &   &  (per cent) & ($\degr$) & (per cent) & ($\degr$) & (per cent) & ($\degr$)   \\
(1)   &     (2)    &   (3)   &   (4)  &  (5)    &   (6)      &   (7)   &   (8)      &   (9)  &  (10)   \\
\hline
 47 & 0.47588896  & 67.411372  & 14.63   & 3.90 $\pm$ 0.61  &   95.7 $\pm$   4.3  &  3.84 $\pm$ 0.46  &   97.2 $\pm$   3.3  &  4.04 $\pm$ 0.41  &  105.9  $\pm$  2.9 \\ 
  59 & 0.50411914  & 67.446995  & 14.88   & 4.50 $\pm$ 0.68  &   89.6 $\pm$   4.3  &  4.58 $\pm$ 0.51  &   98.4 $\pm$   3.1  &  4.64 $\pm$ 0.45  &   97.2  $\pm$  2.7 \\ 
  62 & 0.55995352  & 67.388082  & 13.48   & 4.27 $\pm$ 0.22  &  106.7 $\pm$   1.5  &  4.86 $\pm$ 0.13  &  106.1 $\pm$   0.7  &  4.02 $\pm$ 0.12  &  107.5  $\pm$  0.9 \\ 
  64 & 0.62424757  & 67.411543  & 14.56   & 5.74 $\pm$ 0.38  &  104.3 $\pm$   1.9  &  6.05 $\pm$ 0.20  &  102.7 $\pm$   0.9  &  5.49 $\pm$ 0.18  &  104.4  $\pm$  0.9 \\ 
  92 & 0.52499704  & 67.462054  & 15.46   & 6.03 $\pm$ 0.92  &  103.0 $\pm$   4.3  &  5.15 $\pm$ 0.68  &  101.9 $\pm$   3.7  &  4.51 $\pm$ 0.56  &   99.0  $\pm$  3.5 \\ 
 105 & 0.65972184  & 67.402794  & 13.65   & 6.17 $\pm$ 0.24  &  103.3 $\pm$   1.1  &  5.71 $\pm$ 0.13  &  103.5 $\pm$   0.6  &  4.84 $\pm$ 0.11  &  104.7  $\pm$  0.6 \\ 
 120 & 0.68030330  & 67.416096  & 14.48   & 2.99 $\pm$ 0.37  &  106.7 $\pm$   3.4  &  4.39 $\pm$ 0.20  &  113.0 $\pm$   1.3  &  3.46 $\pm$ 0.19  &  113.7  $\pm$  1.5 \\ 
 133 & 0.40192644  & 67.441032  & 14.94   & 6.31 $\pm$ 0.72  &  105.1 $\pm$   3.2  &  5.59 $\pm$ 0.52  &  106.2 $\pm$   2.6  &  4.44 $\pm$ 0.44  &  106.3  $\pm$  2.8 \\ 
 195 & 0.73841635  & 67.396758  & 14.22   & 8.00 $\pm$ 0.32  &  100.0 $\pm$   1.1  &  7.84 $\pm$ 0.17  &  100.5 $\pm$   0.6  &  6.61 $\pm$ 0.15  &  101.0  $\pm$  0.6 \\ 
 196 & 0.52409579  & 67.340231  & 13.46   & 2.40 $\pm$ 0.22  &   78.1 $\pm$   2.5  &  2.08 $\pm$ 0.12  &   69.5 $\pm$   1.6  &  1.91 $\pm$ 0.12  &   72.7  $\pm$  1.7 \\ 
 198 & 0.70341898  & 67.367529  & 13.47   & 1.88 $\pm$ 0.24  &   65.7 $\pm$   3.5  &  2.18 $\pm$ 0.15  &   79.5 $\pm$   1.9  &  1.76 $\pm$ 0.17  &   75.4  $\pm$  2.5 \\ 
 202 & 0.58241709  & 67.499918  & 14.19   & 5.61 $\pm$ 0.50  &   94.4 $\pm$   2.6  &  5.24 $\pm$ 0.28  &   95.7 $\pm$   1.5  &  4.72 $\pm$ 0.17  &   96.4  $\pm$  1.0 \\ 
 215 & 0.53576488  & 67.333432  & 13.99   & 3.42 $\pm$ 0.30  &   72.8 $\pm$   2.5  &  2.47 $\pm$ 0.18  &   74.6 $\pm$   2.0  &  1.98 $\pm$ 0.19  &   76.1  $\pm$  2.6 \\ 
 240 & 0.71675681  & 67.484199  & 14.61   & 7.49 $\pm$ 0.47  &   99.4 $\pm$   1.8  &  8.23 $\pm$ 0.29  &   95.8 $\pm$   1.0  &  7.27 $\pm$ 0.31  &   94.4  $\pm$  1.2 \\ 
 317 & 0.32552520  & 67.501884  & 12.76   & 5.54 $\pm$ 0.28  &  110.2 $\pm$   1.5  &  5.28 $\pm$ 0.14  &  108.1 $\pm$   0.8  &  4.82 $\pm$ 0.12  &  108.6  $\pm$  0.7 \\ 
 433 & 0.50950597  & 67.569704  & 13.22   & 5.80 $\pm$ 0.26  &  113.3 $\pm$   1.2  &  4.64 $\pm$ 0.16  &  106.5 $\pm$   1.0  &  4.28 $\pm$ 0.18  &  105.7  $\pm$  1.1 \\ 
 503 & 0.23764401  & 67.540179  & 13.55   & 5.37 $\pm$ 0.33  &  102.0 $\pm$   1.7  &  5.27 $\pm$ 0.21  &  103.3 $\pm$   1.1  &  4.93 $\pm$ 0.21  &  103.2  $\pm$  1.2 \\ 
 594 & 0.94921847  & 67.514749  & 11.25   & 1.94 $\pm$ 0.08  &   85.0 $\pm$   1.1  &  1.64 $\pm$ 0.06  &   83.5 $\pm$   1.1  &  1.44 $\pm$ 0.06  &   84.0  $\pm$  1.2 \\ 
 689 & 1.00593420  & 67.502698  & 14.40   & 8.73 $\pm$ 0.35  &   88.2 $\pm$   1.1  &  7.69 $\pm$ 0.23  &   87.8 $\pm$   0.8  &  6.61 $\pm$ 0.17  &   88.9  $\pm$  0.8 \\ 
 742 & 0.71032883  & 67.614514  & 14.87   & 5.88 $\pm$ 0.53  &   93.3 $\pm$   2.5  &  5.44 $\pm$ 0.33  &   95.3 $\pm$   1.7  &  5.43 $\pm$ 0.33  &   97.4  $\pm$  1.7 \\ 
 828 & 1.04825650  & 67.522676  & 12.66   & 8.30 $\pm$ 0.15  &   91.2 $\pm$   0.5  &  8.30 $\pm$ 0.10  &   90.0 $\pm$   0.3  &  7.44 $\pm$ 0.08  &   90.6  $\pm$  0.3 \\ 
1028 & 1.11032740  & 67.533648  & 13.57   & 5.88 $\pm$ 0.23  &  105.8 $\pm$   1.1  &  5.99 $\pm$ 0.15  &  107.0 $\pm$   0.7  &  5.11 $\pm$ 0.12  &  107.5  $\pm$  0.6 \\ 
1106 & 1.19379620  & 67.469519  & 14.35   & 7.66 $\pm$ 0.33  &  100.8 $\pm$   1.2  &  8.07 $\pm$ 0.24  &   98.5 $\pm$   0.8  &  5.47 $\pm$ 0.19  &  102.4  $\pm$  1.0 \\ 
1135 & 1.15739640  & 67.522225  & 14.91   & 8.72 $\pm$ 0.45  &   96.6 $\pm$   1.5  &  7.07 $\pm$ 0.29  &  100.9 $\pm$   1.1  &  6.26 $\pm$ 0.21  &   99.0  $\pm$  1.0 \\ 
1190 & 1.11587840  & 67.561287  & 13.20   & 2.70 $\pm$ 0.20  &   90.3 $\pm$   2.0  &  2.72 $\pm$ 0.15  &   88.2 $\pm$   1.5  &  2.32 $\pm$ 0.13  &   87.2  $\pm$  1.6 \\ 
1273 & 0.70653826  & 67.681432  & 15.07   & 5.66 $\pm$ 0.57  &   93.9 $\pm$   2.8  &  5.61 $\pm$ 0.37  &   94.8 $\pm$   1.8  &  5.03 $\pm$ 0.38  &   97.4  $\pm$  2.1 \\ 
1305 & 0.69299158  & 67.686948  & 14.74   & 5.45 $\pm$ 0.49  &   97.8 $\pm$   2.5  &  4.66 $\pm$ 0.30  &   99.1 $\pm$   1.8  &  4.10 $\pm$ 0.31  &   94.4  $\pm$  2.1 \\ 
1312 & 0.59132421  & 67.692850  & 14.08   & 6.29 $\pm$ 0.37  &  101.1 $\pm$   1.6  &  5.78 $\pm$ 0.20  &  103.8 $\pm$   1.0  &  5.14 $\pm$ 0.17  &  106.1  $\pm$  1.0 \\ 
1451 & 1.17321310  & 67.574163  & 14.12   & 5.26 $\pm$ 0.30  &  104.3 $\pm$   1.6  &  4.82 $\pm$ 0.20  &  108.0 $\pm$   1.1  &  4.30 $\pm$ 0.16  &  109.2  $\pm$  1.0 \\ 
1513 & 0.82238181  & 67.690576  & 12.71   & 2.32 $\pm$ 0.19  &   86.3 $\pm$   2.2  &  2.00 $\pm$ 0.14  &   88.3 $\pm$   1.9  &  1.97 $\pm$ 0.17  &   93.3  $\pm$  2.4 \\ 
1523 & 1.27017620  & 67.506336  & 13.86   & 2.28 $\pm$ 0.27  &   87.9 $\pm$   3.2  &  2.45 $\pm$ 0.20  &   88.2 $\pm$   2.2  &  2.18 $\pm$ 0.17  &   84.5  $\pm$  2.2 \\ 
1569 & 1.22657310  & 67.557681  & 13.60   & 3.17 $\pm$ 0.24  &   88.9 $\pm$   2.0  &  3.19 $\pm$ 0.17  &   86.7 $\pm$   1.5  &  2.67 $\pm$ 0.15  &   88.3  $\pm$  1.6 \\ 
 \hline \hline                                                                                                    
\end{tabular}\\
\hspace{-16cm}
$^\star$From P08
\end{table}
\end{landscape}

\subsection{Distribution of $P_{V}$ and $\theta_{V}$:}
A careful inspection of Fig. \ref{starids_Be59}
reveals two groups of stars characterized by different degree and
direction of polarization.  The first group with relatively small degree of polarization
($\sim$ 2 per cent) and with the orientation nearly parallel to the Galactic disk ($\sim 
86\degr$) may be composed of foreground stars. The stars of this group are randomly distributed
on the plane of the sky. The second group,
whose degree of polarization is significantly higher than the first group and the alignment of
polarization vectors ($\sim$ 102$\degr$) is significantly deviated from the GP, may be composed
of cluster members.  Majority of the stars of the second group are located within the cluster
region ($\sim$ 10 arcmin). 

\begin{figure}
\begin{small}
\resizebox{8.25cm}{8.5cm}{\includegraphics{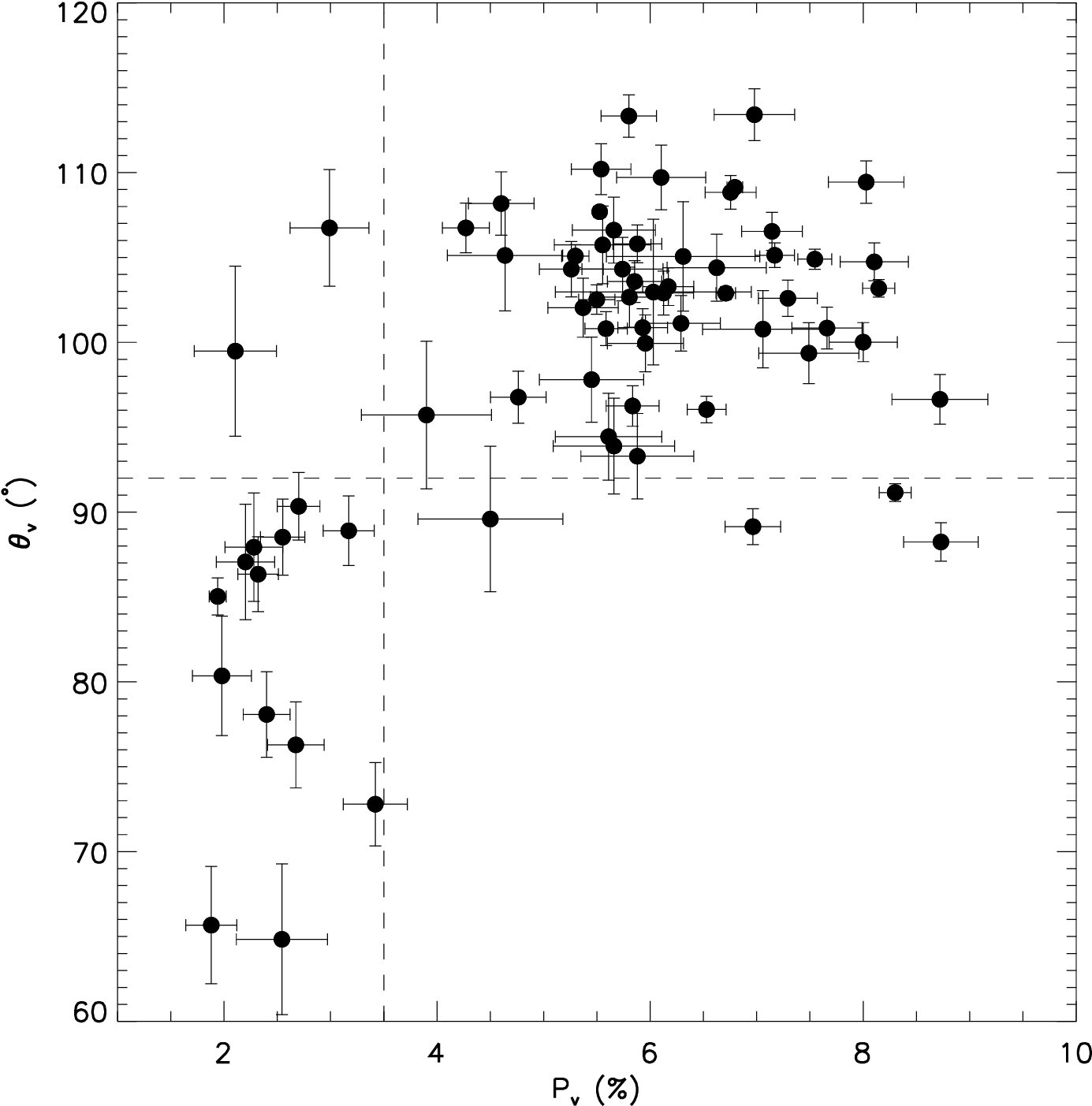}}
\vskip.1cm
\caption{Polarization angle versus polarization in $V$-band for 69 stars towards Be 59. Dashed 
lines are drawn to show the two clearly separated groupings among the observed sample.}
\label{tv_vs_pv}
\end{small}
\end{figure}

Figure \ref{tv_vs_pv} displays the distribution of polarization $P_{V}$ vs polarization angle $\theta_{V} $,
which clearly segregates the above mentioned two groups.
The mean $P_{V}$ and $\theta_{V}$ for the first group is found to be 2.47 $\pm$0.46 per cent
and  81$\degr$$\pm$9$\degr$, respectively.
The mean polarization angle is nearly aligned with the GP, indicating the homogeneity in the
{\it local magnetic field} of the Galaxy towards the direction of Be 59.
The stars belonging to the second group are more
polarized ($P_{V}\ga4.0$ per cent) and having $\theta_{V}\ga90\degr$. The mean polarization angle of
the second group (103$\degr$$\pm$5$\degr$) significantly deviates from the GP.
The magnetic field associated with the parental molecular cloud may have been perturbed
during the cloud collapse or due to the strong stellar winds or
supernova explosions (e.g., Waldhausen, Mart\'{i}nez \& Feinstein 1999).
The stars belonging to the second group could be either the
cluster members or background stars. The second group of stars show large dispersion in polarization
but a very small dispersion in polarization angle. The large dispersion in $P_{V}$ ($4 - 8.7$ per cent)
could be attributed to the differential reddening within the cluster. Similar type of segregated groups have already
been reported in a few cases, namely Trumpler 27 (Feinstein et al. 2000), Hogg 22 \& NGC 6204 (Mart\'{i}nez et al. 2004),
NGC 5749 (Vergne et al. 2007) and NGC 6250 (Feinstein et al. 2008).

\begin{figure}
\begin{small}
\resizebox{8.5cm}{10.5cm}{\includegraphics{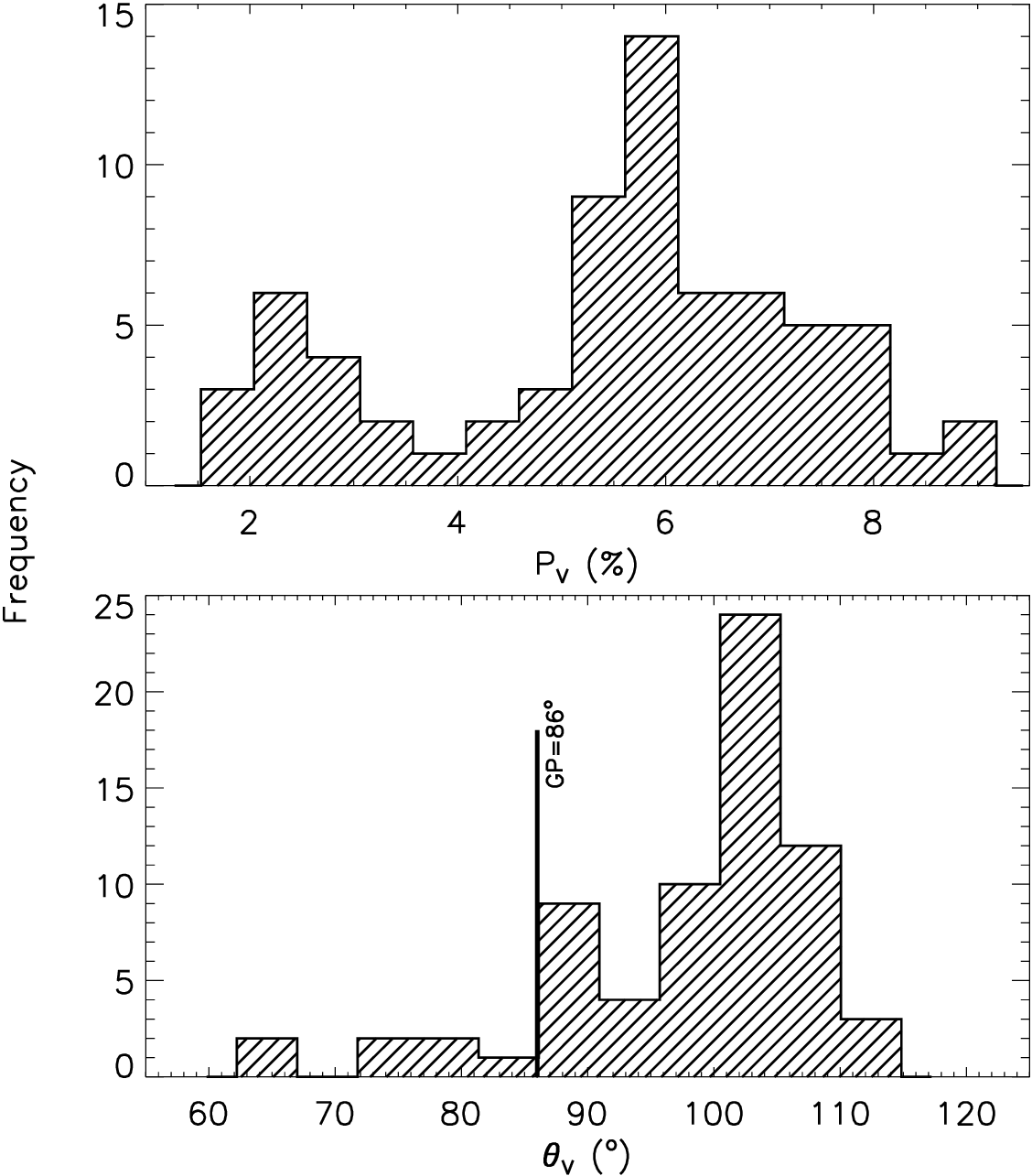}}
\vskip.1cm
\caption{Histograms for $P_V$ (upper panel) and $\theta_V$ (lower panel) for 69 stars towards Be 59.
Thick line drawn in the lower panel represents the projection of the Galactic plane at 86$\degr$.}
\label{Distri_Pv_and_Tv}
\end{small}
\end{figure}

The distribution of $P_{V}$ for all the observed 69 stars is shown in the upper panel 
of Fig. \ref{Distri_Pv_and_Tv}, which clearly reveals two separate distributions for 
field and probable cluster members. The distribution for field stars 
peaks at $\sim$ 2 per cent, whereas that for the probable members peaks at  5.5 per cent with an  
extended tail towards higher polarization, which could be either due to highly 
extincted probable members, background stars or due to the presence of different 
populations of dust grains with different polarizing properties. 
The lower panel of Fig. \ref{Distri_Pv_and_Tv} shows 
the distribution of $\theta_{V}$, which reveals that the distribution of 
probable cluster members lies in the range of 95$\degr$$-$115$\degr$ 
with a peak at $\sim$105$\degr$. 

To have a better understanding of the nature of the dust component and the magnetic field 
associated with the foreground and intra-cluster medium,  it is essential to find 
out the members associated with the cluster versus the stars located in the 
foreground/background of the cluster. In our previous study (E11) we have shown that 
polarimetry in combination with the $(U-B)-(B-V)$ two-colour diagram (TCD) 
can yield a better identification of probable members than photometry alone. 
In the ensuing section we will discuss the determination of membership using the polarization 
properties in combination with $(U-B)-(B-V)$ TCD.

\subsection{Member identification}\label{member_identi}

\begin{figure*}
\begin{small}
\resizebox{16cm}{16cm}{\includegraphics{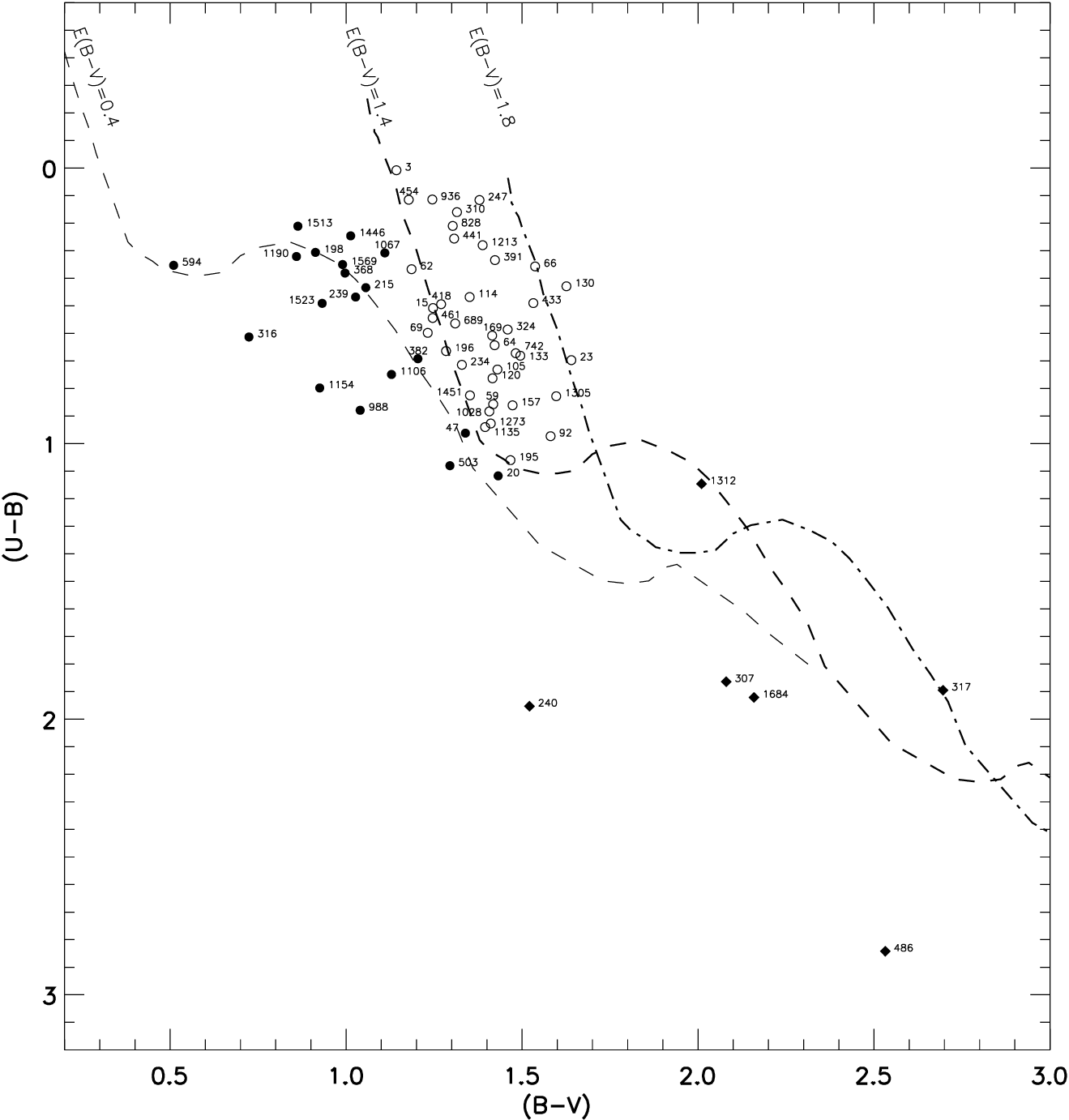}}
\vskip.1cm
\caption{$(U-B)$ versus $(B-V)$ TCD.
Theoretical ZAMS taken from Schmidt-Kaler (1982) is shifted along a reddening vector with an adopted slope 
of $E(U-B)$/$E(B-V)$=0.72, to the match
the observed colours. The probable members and field stars are shown by open and filled circles. The late type 
stars are shown with filled diamond symbols.}
\label{UBBVCCD}
\end{small}
\end{figure*}

\begin{figure*}
\begin{small}
\resizebox{16cm}{16cm}{\includegraphics{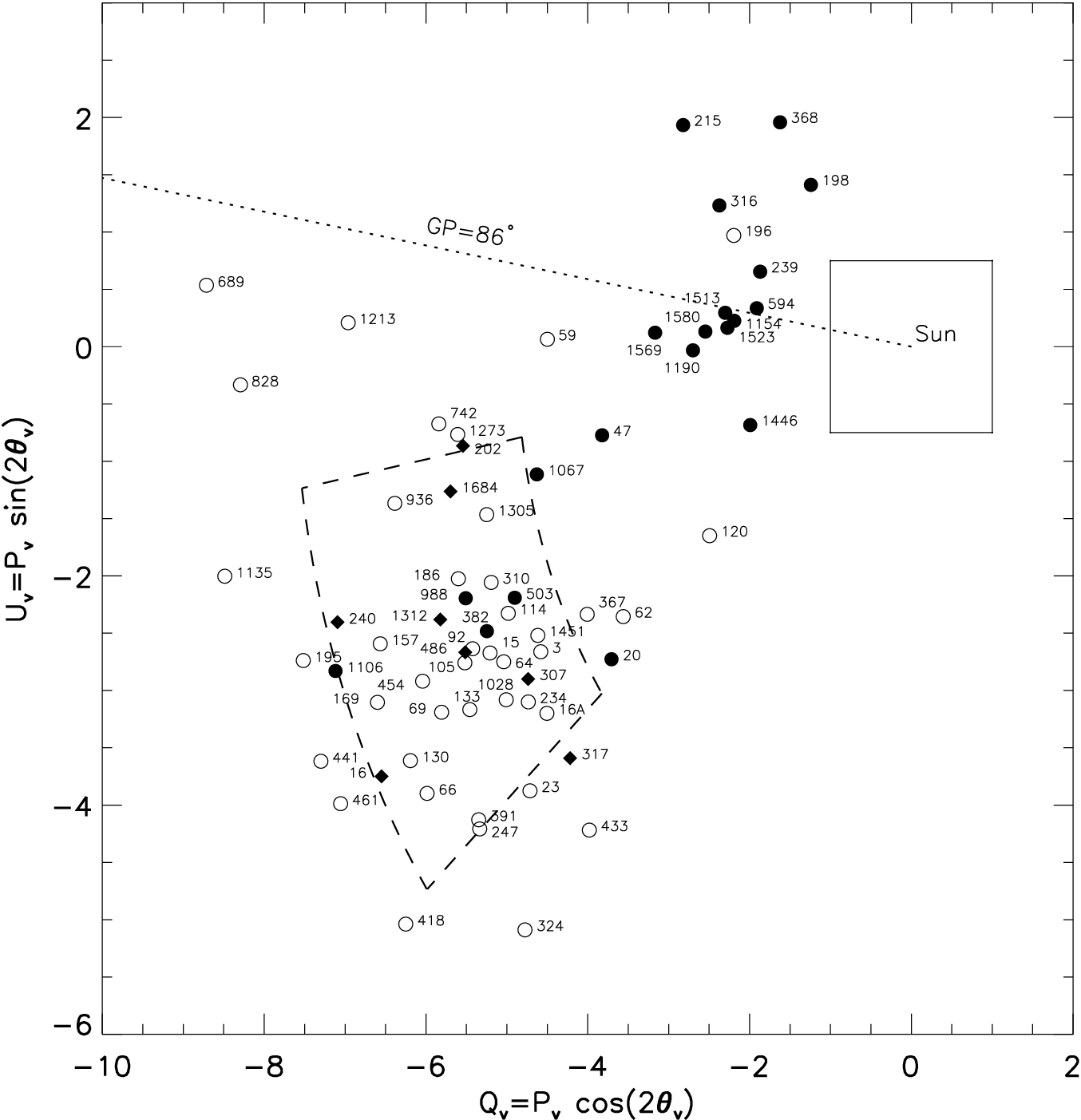}}
\vskip.1cm
\caption{$Q_V$ versus $U_V$ of 69 stars. Symbols are same as that of Fig. \ref{UBBVCCD}. GP is also drawn 
with dotted line. The square covered by $Q_V$=0 and $U_V$=0 is the dust less Solar neighborhood. $1\sigma$ box 
is drawn with dashed line using the mean and standard deviation of $P_V \pm \sigma_{P_{V}}$=6.22$\pm$1.21 per cent and 
$\theta_V \pm \sigma_{\theta_{V}}$=102$\pm$6$\degr$.}
\label{stokes_plane}
\end{small}
\end{figure*}

$(U-B)-(B-V)$ TCD is one of the useful tools to identify probable members of a cluster.
It is expected that all member stars have $E(B-V)$ values comparable to the mean $E(B-V)$ 
value of the cluster since the cluster stars have formed out of the same molecular cloud 
and consequently have the same distance and age. In comparison, the field population is expected to be 
less or highly extincted depending on whether they are foreground or background 
to the cluster. Fig. \ref{UBBVCCD} shows the $(U-B)-(B-V)$ TCD for only 
63 stars as the $U-B$ colours are not available 
for 6 stars. In Fig. \ref{UBBVCCD} the zero-age-main-sequence (ZAMS) 
from Schmidt-Kaler (1982) is shifted along 
a normal reddening vector having a slope of $E(U-B)/E(B-V)$ = 0.72. 
The TCD shows a variable reddening in the cluster region 
with $E(B-V)_{min}$$\sim$1.4 mag and $E(B-V)_{max}$$\sim$1.8 mag. The cluster members (shown by
open circles) seem to have spectral types earlier than $A0$.
The TCD manifests the presence of a foreground population shown by filled circles. 
The foreground population reddened by $E(B-V)$=0.4 mag is found to be located at $\sim$ 470 pc (P08). 
The other field star (late type) population is shown by filled diamonds. 

The polarimetric observations can be used as a tool to identify member stars in a Galactic open cluster, 
particularly, when the field stars have colours similar to those 
of cluster members (Mart\'{i}nez et al. 2004; Vergne et al. 2007; Feinstein et al. 2008; Vergne et al. 
2010; Orsatti et al. 2010; E11). The individual Stokes parameters of the polarization 
vector of the $V$-band, $P_V$, given by $Q_{V}=P_{V}\cos(2\theta_{V})$ and 
$U_{V}=P_{V}\sin(2\theta_{V})$ are estimated for all the observed stars toward Be 59 and presented 
on a $U_{V}$ versus $Q_{V}$ plot, known as the Stokes plane, in Fig. \ref{stokes_plane}. 

The measured degree of polarization of a star depends on the cumulative amount of aligned dust grains that lie 
along the line of sight, and hence the degree of polarization would be similar, lower or higher depending 
on whether it is member, foreground or background to the cluster. Likewise, the position angles 
of the cluster members would be similar, but different for foreground or background field stars 
as light from them could have contributions from different or additional dust 
components. Hence the cluster members are expected to show a grouping in the $U_V-Q_V$ plane, 
while non-members are expected to show a scattered distribution. 
Therefore $U_V-Q_V$ plot could be a useful tool to identify the members of a cluster. 
The stars with intrinsic polarization (due to asymmetric distribution of matter 
around young stellar objects (YSOs)) and/or rotation in their polarization angles may also create 
scattered distribution in the $U_V-Q_V$ plane. Nebulous background in case of star-forming 
regions would also create small intrinsic polarization and hence possibly the scattered distribution.

Fig. \ref{stokes_plane} shows two prominent groupings 
around $U_V \sim -2.5$ and $Q_V \sim -5.5$ (first group) and $U_V \sim 1.0$ and 
$Q_V \sim -2.0$ (second group). 
The grouping at $U_V \sim -2.5$ and $Q_V \sim -5.5$ should be due to the cluster members.
Fifty percent (13 out of 27) of the field stars, identified on the basis 
of $(U-B)-(B-V)$ TCD (Fig. \ref{UBBVCCD}), can be noticed mainly 
around $U_V \sim 1.0$ and $Q_V \sim -2.0$. The remainder of the probable field stars show a 
scattered distribution in $U_V-Q_V$ plot; however some of the probable field stars 
are found to mingle with the probable cluster members. 
To further elucidate the membership we plot a box with 
dashed line in $U_{V}-Q_{V}$ plot having boundaries of mean $P_{V}\pm \sigma_{P_{V}}$ (6.22$\pm$1.21 per 
cent) and mean $\theta_{V} \pm \sigma_{\theta_{V}}$ (102$\pm$ 6$\degr$) obtained using the probable member 
stars (open circles) identified in Fig \ref{UBBVCCD}. The stars shown with open circles and lying within the 
1$\sigma$ box of the mean $P_{V}$ and $\theta_{V}$ could be probable members of the cluster. 
It is apparent from Fig. \ref{UBBVCCD}, i.e., the $(U-B)-(B-V)$ TCD that the 
majority of the stars located within the $1\sigma$ box follow the general reddening of the cluster 
region, hence are probable members of the cluster.

Stars \# 59, 120, 689, 828 and 1213 are located significantly away from the $1\sigma$ box. 
The stars \#59 ($P_V$=4.5 per cent, $\theta_V$=90$\degr$) and 
\#120 ($P_V$=3.0 per cent, $\theta_V$=107$\degr$), even though 
located spatially within the cluster region (see Fig. \ref{starids_Be59}) 
and with photometric colours consistent with membership, 
their $P_V$ value or $\theta_V$ value is not comparable to the cluster region 
(see Fig. \ref{stokes_plane}), so are considered as non-members.  
The stars \# 689, 828 and 1213 are located outside the cluster region. 
These stars show relatively large polarization ($\sim$ 6.6 to 8.7 per cent) but their $\theta_V$ values 
($\sim$ 88$\degr$ to 91$\degr$) are significantly different from those of probable cluster members 
(Tables \ref{BVRI_pol_data} and \ref{VRI_pol_data}) and are comparable to the GP. 
Hence these are considered as non-members. The stars \# 23, 195, 247, 324, 391, 418, 441, 433 and  461 are 
distributed outside but near the boundary of  $1\sigma$ box. Of these 9 stars (barring \# 23, 195 and 
324) have relatively higher value of $\overline\epsilon$ 
(cf. Fig. \ref{raddistri_eps_and_eps_vs_pmax}, Sec \ref{dustproperties}) indicating 
rotation in their polarization angles. These stars have $E(B-V)$ in the 
range of 1.4$-$1.8 mag (cf. Fig. \ref{UBBVCCD}), so could be members of the cluster. 
Three stars, namely \# 20, 62 and 367 are also located near the 
boundary of the $1\sigma$ box. The $P_V$ values of these stars are in the range 
of 4.3 per cent to 4.6 per cent which is the lower limit of $P_V$ value for the 
cluster member stars (see Tables \ref{BVRI_pol_data} and \ref{VRI_pol_data}). 
The $\theta_V$ values range from 105$\degr$$-$108$\degr$. These could also be 
members of the cluster. 

There seems to be a less prominent grouping around $U_V \sim -1.0$ and $Q_V \sim -5.0$ consisting 
of stars \# 202, 742, 936, 1067, 1273, 1305 and 1684 (cf. Fig. \ref{stokes_plane}). 
We refer to this group as the third group. 
Interestingly, all these stars (except 202) are located spatially at the same region towards the Northern part of the
cluster Be 59. This group of stars is located near the edge of 1$\sigma$ box. 
The mean value of $P_V$ (5.68$\pm$0.53 per cent) is comparable to the $P_V$ values 
of the cluster region. However, the mean $\theta_V$ (=$95 \degr \pm 2\degr$) is significantly different 
from that in the cluster region. These stars are considered as field stars. 

The colours of stars \# 1028, 1135 and 1451 are comparable to those of the cluster members and lie in the 
$E(B-V)$$\simeq$1.4 to 1.8 mag range (presuming that these stars have spectral types earlier than $A0$). 
Star \# 1135 has $P_V$ (8.12$\pm$0.45 per cent) higher than those for cluster members, however 
$\theta_V$ (96.6$\pm$1.5$\degr$) is smaller than the value for cluster stars. This star is located 
outside the 1$\sigma$ box and also located outside the estimated boundary of 
the cluster (Fig. \ref{starids_Be59}), 
hence this star is considered as a non-member. Stars \# 1028 and 1451 have $P_V$ and $\theta_V$ 
values both comparable to those of the cluster members but located outer side boundary of the cluster, 
hence the membership of these stars is uncertain. 

Star \# 196 is located near the solar neighborhood in the $Q_V$ and $U_V$ diagram. Its colours are 
consistent with those of early type members of the cluster,  which yields $E(B-V)$=1.4 mag, but its 
location in the $Q_V-U_V$ diagram (Fig. \ref{stokes_plane}) manifests that it should 
be a field star, hence this star is considered as a field star 
for further discussion. The star \# 1106 is located near the 1$\sigma$ box boundary 
(cf. Fig. \ref{stokes_plane}). On the basis of its $(U-B)-(B-V)$ colours and its location outside the 
cluster boundary, we consider it as a non-member. 

The stars \# 382, 503 and 988 are well separated from the cluster probable members in the $(U-B)$-$(B-V)$ TCD 
(cf. Fig. \ref{UBBVCCD}), however they lie  within 1$\sigma$ box in the $Q_V-U_V$ plot (Fig. \ref{stokes_plane}). 
Since stars \# 382 and 503 lie within the boundary of the cluster and $P_V$ and $\theta_V$ values are comparable to the 
cluster stars, these are considered as probable members. 
Star \# 988 lies outside the boundary of the cluster, its membership is uncertain.

The probable members of the cluster identified using $U_V-Q_V$ and colour-colour diagrams are given in Table 
\ref{ebv_memb_results}. The member, probable member and non-member stars are represented with "M", "PM" 
and "NM" respectively. Stars with uncertainty in their membership determination are indicated with a "?" symbol. 

Since cluster members seem to have spectral types earlier than $A0$ (see Fig. \ref{UBBVCCD}), 
the reddening $E(B-V)$ for the member stars has been estimated using the $Q$-method (Johnson \& Morgan 1953). 
As seen in Fig. \ref{UBBVCCD}, all the field stars have spectral types later than $A0$. The reddening $E(B-V)$ 
for field stars is estimated visually using the slide-fit method of ZAMS along the reddening vector. 
The estimated values of $E(B-V)$ are given in Table \ref{ebv_memb_results}. 
The colours of stars \# 828 and 1213 
suggest spectral types earlier than $A0$. The colours of 
star \# 689 suggest two reddening values, ($E(B-V)$=0.70 mag for a spectral type 
later than $A0$, or $E(B-V)$=1.44 mag for a spectral type earlier than $A0$). 
Star \# 1135 has $P_V$=8.72 per cent, which is comparable to the cluster members, however its $\theta_V$ value 
(96.6$\degr$) is different from the cluster region. Its colours also suggest two values of $E(B-V)$ (1.17 mag for 
a spectral type later than $A0$, or 1.42 mag for a spectral type earlier than $A0$). 
Similarly, colours for stars \# 1028, 1106, and 1451 have two possible values of reddening. 
These are mentioned in Table \ref{ebv_memb_results}. 

The colours of the third group of stars (\# 742, 936, 1067, 1273, 1305 and 1684) 
suggest that their $E(B-V)$ values are 
in the range of $\sim$ 0.95$-$1.30 mag. The location of the star \# 936 in the $(U-B)-(B-V)$ TCD suggests 
an O/B spectral type. The mean values of $P_V$ and $\theta_V$ for these stars are 
5.69 $\pm$ 0.58 per cent and 96$\degr$$\pm$2$\degr$ respectively. 
This group of stars may lie between the foreground 
stars ($\sim$ 470 pc) and the cluster. 
Assuming a mean $E(B-V)$=1.20 mag, 
the $V_{0}/(V-I)_{0}$ CMD indicates that
these stars are distributed at a distance of $\sim$ 700 pc. 

\begin{table*}
\caption{The $E(B-V)$  values estimated using the $(U-B)-(B-V)$ TCD. The membership information is 
also mentioned against each star ID}
\label{ebv_memb_results}
\begin{tabular}{cccccccc} \hline  \hline
\multicolumn{3}{|c|}{stars with $BV(RI)_{C}$ pass-band data} &\multicolumn{2}{|c|}{} & \multicolumn{3}{|c|}{stars with $V(RI)_{C}$ pass-band data}\\
\hline
Star ID $^\star$ &  $E(B-V)$ &  membership & &  & Star ID $^\star$ & $E(B-V)$ & membership \\
                 &  (mag)    &            & & &      &  (mag)  &             \\
(1)              &   (2)     &   (3)      & & &   (4)    &  (5)    &  (6)        \\
\hline \hline
     3 & 1.41 &   M                    & & &     47 & 0.43$^\dagger$, 1.20  & NM  \\                 
    15 & 1.38 &   M                    & & &     59 & 0.63$^\dagger$, 1.13$^\dagger$  & NM \\                 
    16 & -    &   NM                   & & &     62 & 1.35    & M \\                       
   16A & -    &   M                    & & &     64 & 1.55  &  M \\                         
    20 & 0.48$^\dagger$, 1.40$^\dagger$  & M    & & &     92 & 1.64 &   M \\                         
    23 & 1.80 &   M                    & & &    105 & 1.53 &  M \\                          
    66 & 1.79 &   M                    & & &    120 & 0.70$^\dagger$, 1.06$^\dagger$  & NM  \\                
    69 & 1.33 &   M                    & & &    133 & 1.63 &  M \\                          
   114 & 1.52 &   M                    & & &    195 & 1.47  &  M \\                         
   130 & 1.87 &   M                    & & &    196 & 0.57$^\dagger$ &  NM  \\                       
   157 & 1.54 &   M                    & & &    198 & 0.40$^\dagger$ &  NM \\                        
   169 & 1.55 &   M                    & & &    202 &  -   &  NM \\                         
   186 &  -   &   M                    & & &    215 & 0.63$^\dagger$ &  NM  \\                       
   234 & 1.41 &   M                    & & &    240 &  -   &   NM \\                        
   239 & 0.33$^\dagger$, 0.65$^\dagger$ & NM             & & &    317 & 1.81$^\dagger$  &  NM  \\                      
   247 & 1.67 &  M                     & & &    433 & 1.74 &   M  \\                        
   307 & 0.95$^\dagger$   & NM        & & &   503 & 1.30   & PM \\                        
   310 & 1.58 &   M                    & & &    594 & 0.35$^\dagger$, 0.51   & NM \\                 
   316 & 0.65$^\dagger$    & NM                 & & &    689 & 0.70$^\dagger$, 1.44   & NM \\                 
   324 & 1.61  & M                     & & &    742 & 0.95$\dagger$, 1.61  & NM \\        
   367 &  -    & M                     & & &    828 & 1.54  & NM \\                         
   368 & 0.55$^\dagger$  &  NM                  & & &   1028 & 1.14$^\dagger$, 1.45   & ? \\                  
   382 & 1.26  & PM                    & & &   1106 & 1.17   & NM   \\                      
   391 & 1.65  & M                     & & &   1135 & 1.17$^\dagger$, 1.42  & NM  \\                 
   418 & 1.41  & M                     & & &   1190 & 0.46$^\dagger$  & NM  \\                       
   441 & 1.53  & M                     & & &   1273 & 0.56$^\dagger$, 1.16$^\dagger$ & NM  \\                 
   454 & 1.42  & M                     & & &   1305 & 0.95$^\dagger$, 1.17$^\dagger$ & NM  \\                 
   461 & 1.36  & M                     & & &   1312 & 1.34$^\dagger$ & NM \\                         
   486 &  -    & NM                    & & &   1451 & 1.08$^\dagger$, 1.40  & ?  \\                  
   936 & 1.50 &  ?                     & & &   1513 & 0.30$^\dagger$    & NM   \\                    
   988 & 1.05$^\dagger$  & ?           & & &   1523 & 0.63$^\dagger$    & NM  \\                     
  1067 & 0.55$^\dagger$, 1.27  & NM   & & &   1569 & 0.45$^\dagger$    & NM \\ 
  1154 & 0.85$^\dagger$   & NM        & & & & & \\
  1213 & 1.63 &  NM                    & & & & & \\
  1446 & 0.45$^\dagger$  & NM                   & & & & & \\
  1580 &  -     & NM                   & & & & & \\
  1684 & 0.95$^\dagger$ NM            & & & & & \\
\hline \hline                                                                                                    
\end{tabular}\\
$^\star:$ From P08\\
$^\dagger$: $E(B-V)$ values were obtained using slide and fit method\\                             
M: Members\\
NM: Non-members\\
PM: Probable members \\
$?$: Stars with uncertainty in their membership\\
\end{table*}

\begin{figure}
\begin{small}
\resizebox{8.5cm}{8.5cm}{\includegraphics{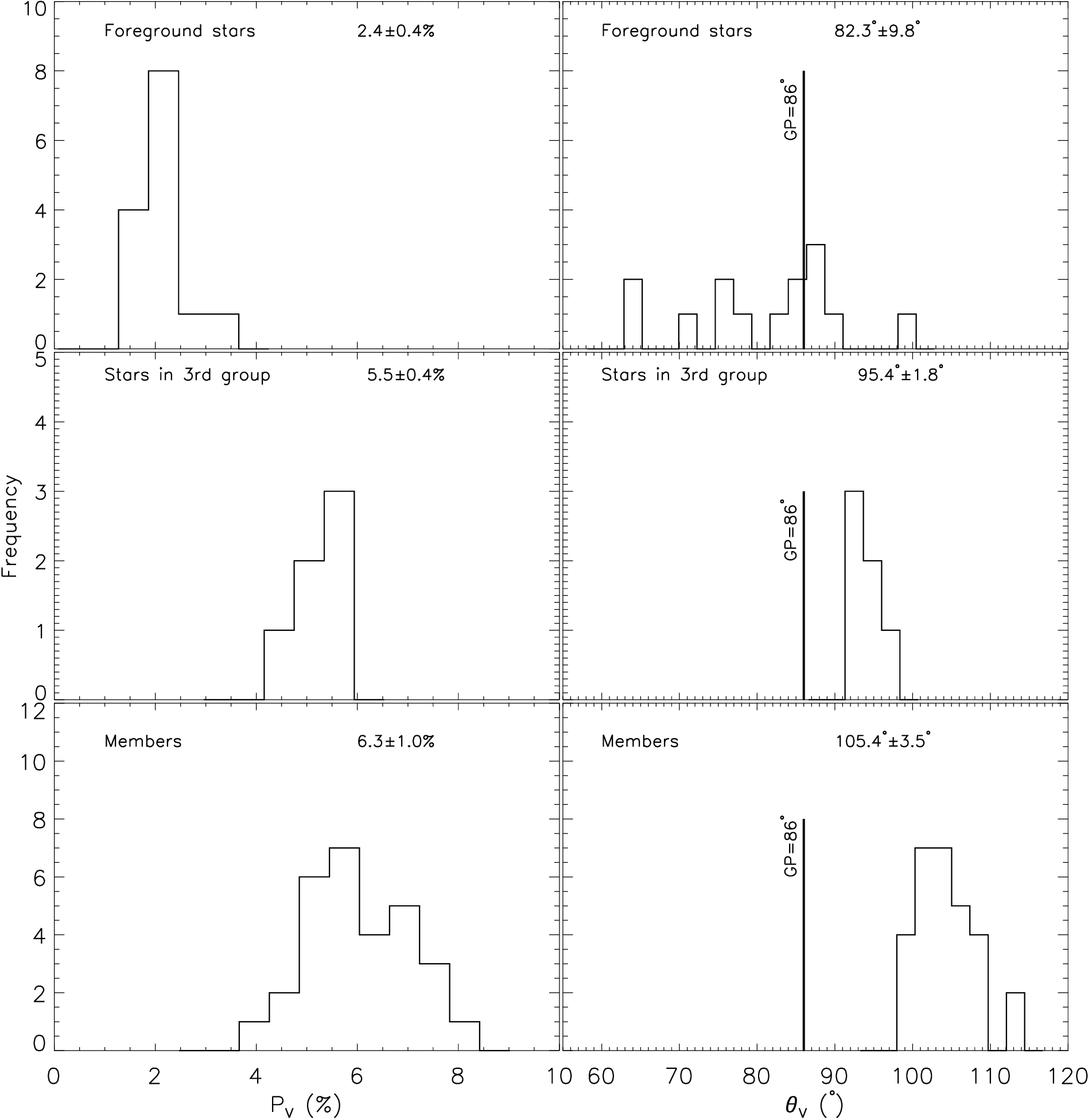}}
\vskip.1cm
\caption{Histograms of $P_V$ (left panels) and $\theta_{V}$ (right panels) for the foreground stars, 
stars in 3rd group and the cluster members. For comparison, the GP is also drawn with thick line at 86$\degr$. 
The mean and standard deviation of $P_V$ and $\theta_{V}$ values for each group are also mentioned.}
\label{PT_M_NM_CONF}
\end{small}
\end{figure}

\begin{table*}
\centering
\caption{The estimated mean values of $P_V$, $\theta_V$, $Q_V$ and $U_V$ for the foreground stars, 
the stars in 3rd group and the cluster members. The distance information is also mentioned}
\label{mean_pv_tv_m_nm_2ndgrp}
\begin{tabular}{ccccccc}\hline \hline
     &  $\overline{P_V}$ (per cent)    & $\overline{\theta_V}$ ($\degr$)  & $\overline{Q_V}$ & $\overline{U_V}$ & No. of stars & Distance (pc)  \\
\hline
First group     &  2.44 $\pm$ 0.45 & 82$\pm$10 & $Q_{f}$=-2.35 & $U_{f}$=0.65 & 14 & 470 \\
Third group     &  5.55 $\pm$ 0.41 & 95$\pm$2 & $Q_{3}$=-5.43 & $U_{3}=$ -1.04 & 6 & 700 \\
Member stars    &  6.34 $\pm$ 1.05 & 105$\pm$4 & $Q_{m}$=-5.45 &$U_{m}$ -3.24 & 29 & 1000 \\
\hline
\end{tabular} \\
\end{table*}


Fig. \ref{PT_M_NM_CONF} shows the distribution of $P_V$ (left panels) and $\theta_V$ (right panels) 
for the identified foreground stars (first group and third group of stars) and cluster members. 
The mean and standard deviation 
of $P_V$ and $\theta_V$ of these three groups of stars are given in a Table 
\ref{mean_pv_tv_m_nm_2ndgrp} and also shown in Fig. \ref{PT_M_NM_CONF}. 
The mean values of $Q_V$ and $U_V$ are also listed in Table \ref{mean_pv_tv_m_nm_2ndgrp}. 
Fig. \ref{PT_M_NM_CONF} shows the mean degree of polarization, polarization angle as well as the deviation 
of the mean polarization angle from the GP increases systematically with increasing distance. 

\begin{figure}
\begin{small}
\resizebox{8.5cm}{9.0cm}{\includegraphics{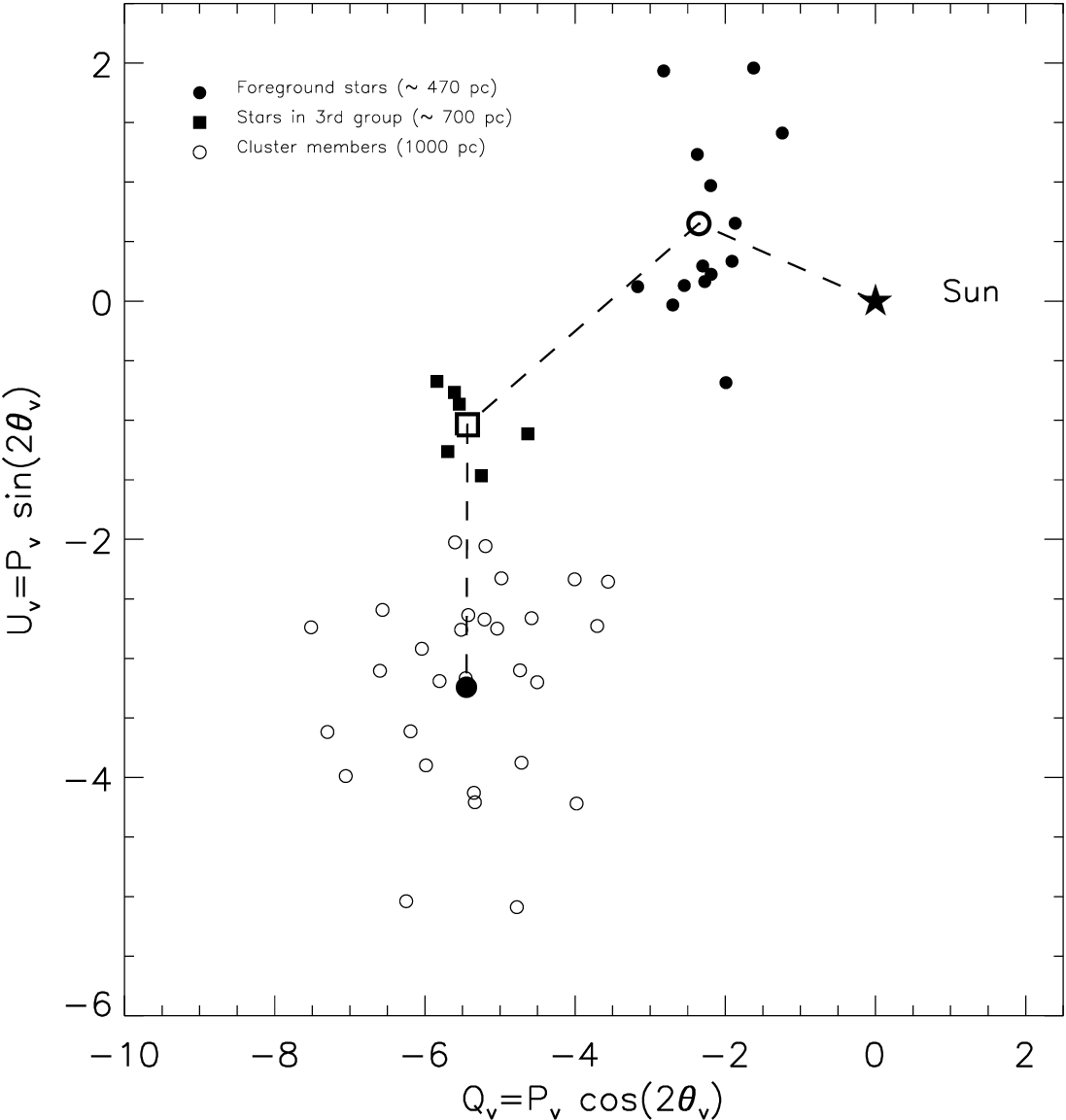}}
\vskip.1cm
\caption{$U_V$ versus $Q_V$ diagram using the stars with known membership. The stars near the Sun, 3rd group of stars and
the cluster members are shown with filled circles, filled squares and open circles respectively.
The mean Stokes parameters for these three groups of stars are also shown with bigger open circle, open square and filled circle 
symbols. The filled star symbol denotes the position of Sun at $Q_V$=0 and $U_V$=0}
\label{stokes_plane_M_NM_2ndgrp}
\end{small}
\end{figure}

\subsection{Dust distribution}
\label{dust_distri}

The stokes plane can be used effectively not only to determine the membership but also to 
delineate the variations in the interstellar environments, e.g., the distribution of dust layers, 
the role of dust layers in polarization, the associated magnetic field orientation etc. 
The vector that connects two points in the Stokes plane represents the amount of polarization, while 
any change in the direction of vectors is related to change in the polarization angle as 
seen by us toward a particular line of sight. 
If the dust grains are oriented/aligned uniformly (i.e., uniform magnetic field orientation) 
then the degree of polarization is expected to increase with 
distance but the direction of polarization (polarization angle) should remain the same and hence the 
Stokes vector should not change its direction with increasing distance. 
For example, in the case of NGC1893 (E11) the degree of polarization 
was found to increase with distance whereas the direction of polarization remains almost constant (cf. their Fig. 5). 

To understand the dust distribution towards Be 59, we compare the polarization measurements 
of foreground stars, the stars of third group, and the cluster members as shown in 
Fig. \ref{stokes_plane_M_NM_2ndgrp}. 
The mean Stokes parameters of these three groups of 
stars are also marked with large open circle, open square and filled circle respectively 
and are connected with dashed lines. 

\begin{table*}
\centering
\caption{The estimated net mean values of $P_V$, $\theta_V$, $Q_V$ and $U_V$ due to the dust layers in front 
of first and third group of stars as well as due to the intra-cluster medium.}
\label{dust_layer_properties}
\begin{tabular}{ccccc}\hline \hline
Dust layers & $\overline{P_V}$ (per cent)    & $\overline{\theta_V}$ ($\degr$)  & $\overline{Q_V}$ & $\overline{U_V}$  \\
\hline
In front of first group of stars     &  2.44 & 82 & $Q_{f}$=-2.35 & $U_{f}$=0.65 \\
In front of third group of stars     &  3.51 & 104 & $Q_{3}-Q_{f}$=-3.08  & $U_{3}-U_{f}$= -1.69  \\
Intra-cluster medium  &  2.20 & 135 & $Q_{m}-Q_{3}$=-0.02 & $U_{m}-U_{3}$=-2.20 \\
\hline
\end{tabular} \\
\end{table*}

The individual polarization properties of the dust layers have been estimated by subtracting their 
foreground contribution and the same are given in Table \ref{dust_layer_properties}. 
The polarization measurements of first group of foreground stars (mean $P_V$=2.44$\pm$0.45 per cent and mean 
$\theta_V$=82$\pm$10$\degr$) located at $\sim$ 470 pc indicate the presence of a dust layer at $\la$ 470 pc. 
The orientation of magnetic field of this layer is found to be comparable to the GP. 
The third group of stars, lying between the first group and the cluster Be 59, show different 
polarization measurements ($P_V$=5.55$\pm$0.41 per cent and $\theta_V$=95$\degr$$\pm$2$\degr$) from those of 
first group or the cluster members. This fact manifests that there must be another dust layer at the 
distance $\ga$500 pc that polarizes the star light of the second group of stars by $\sim$ 3.5 per cent 
(see Table \ref{dust_layer_properties}). 
The dust grains in this dust layer are found to be
aligned significantly differently ($\sim$ 104$\degr$, see Table \ref{dust_layer_properties}) from that 
in the first dust layer ($\sim$ 82$\degr$). The polarization measurements for cluster members 
($P_V$=6.34$\pm$1.05 per cent and $\theta_{V}$= 105$\pm$4$\degr$) also indicate different polarization 
properties of intra-cluster dust. Table \ref{dust_layer_properties} suggests that intra-cluster medium 
also polarizes the cluster members by $\sim$ 2.20 per cent. 
The dust grains of the intra-cluster material are aligned significantly differently ($\sim$135$\degr$) 
from those in the two foreground dust layers. 
The polarization angles found to increase with increase 
in distance from the Sun (cf. Table \ref{dust_layer_properties} and Fig. \ref{stokes_plane_M_NM_2ndgrp}). 
This systematic change in the alignment of dust grains may cause {\it depolarization} 
(less polarization efficiency). 
This issue will be discussed in the ensuing section. 

\begin{figure}
\centering
\resizebox{8.5cm}{13.5cm}{\includegraphics{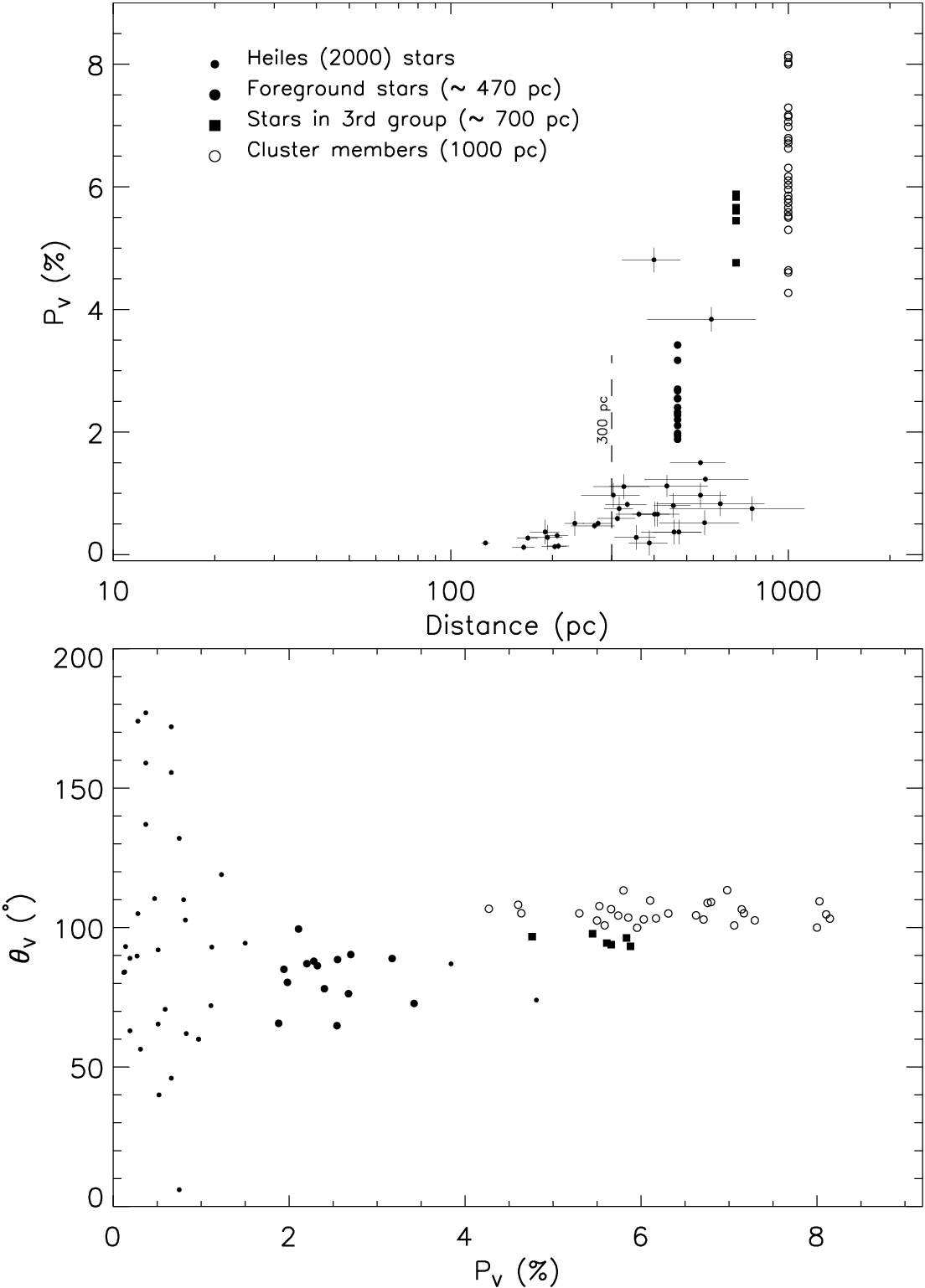}}
\vskip.1cm
\caption{Distance versus polarization (upper panel) for the stars in the direction to Be 59. 
Polarization and distance information for the stars located at distance $\la$ 500 pc is 
obtained from Heiles (2000) and van Leeuwen (2007) respectively. Polarization versus polarization angle (lower panel).} 
\label{dist_p_t_hh}
\end{figure}

To further study the dust distribution for distance $\la$ 500 pc towards the direction of 
Be 59, we select stars with polarization measurements from the catalogue by Heiles (2000) and having 
Hipparcos parallax measurements (van Leeuwen, 2007) in a 10$\degr$ radius around Be 59. 
We selected stars by applying the following 
criteria: (a) $P_V$$\textgreater$ 0.1 per cent, 
(b) ratio of parallax error to the parallax, i.e., $\sigma_{\pi_{H}}/\pi_{H}$ $\la$ 0.5 and 
(c) stars without having any emission features or 
photometric variability (with the help of SIMBAD). 
Figure \ref{dist_p_t_hh} shows polarization versus distance (upper panel) 
and polarization versus polarization angle (lower panel) for the stars studied 
in the present work (same symbols as in Fig. \ref{stokes_plane_M_NM_2ndgrp}) 
as well as the stars (shown with dots) from Heiles (2000). 
Figure \ref{dist_p_t_hh} indicates a sudden increase in the polarization at $\sim$300 pc, $\sim$500 pc 
and $\sim$700 pc, which suggests the presence of three dust layers at $\sim$ 300 pc, $\sim$500 pc and $\sim$ 700 pc 
towards Be 59. 

As shown in the lower panel, the polarization angles of the Heiles stars (dots) are 
distributed randomly, which indicates that the magnetic field orientation in the nearby but in an extended 
region towards the direction of Be 59 is not as organized as compared with that in the cluster region. 
Hence the magnetic field in the intra-cluster medium seems to be more confined. 

Neckel \& Klare (1980) have studied the reddening distribution in the GP 
with $|b|$$\la$ 7.6$\degr$ using the extinction and distances computed for 
individual stars. The $A_V$ map towards the direction of Be 59 by Neckel \& 
Klare (1980) [see their Figure 6a, 4 (115/3)] shows an increase in $A_V$ by $\sim$ 0.6 mag at 
the distance of $\sim$ 300 pc, indicating the presence of a dust layer at this distance. 
For a normal reddening law, $A_V$ $\sim$ 0.6 corresponds to $E(B-V)$ $\simeq$ 0.20 mag. 
This value of $E(B-V)$ yields a polarization of $\sim$ 1 per cent ($P$=5$\times$$E(B-V)$) which is 
in accordance with the dust layer at $\sim$ 300 pc as shown in Fig. \ref{dist_p_t_hh}.
At $\simeq$ 800 pc the $A_V$ further increases and reaches $\sim$ 1 mag. 
The $A_V$ has a steep rise after 800 pc and at a distance of 1 kpc the $A_V$ is $\sim$ 3 mag corresponding 
to the $E(B-V)$ of $\sim$ 1 mag, which is consistent with the cluster's foreground 
reddening. The present polarimetric results are therefore consistent with the reddening distribution given by 
Neckel \& Klare (1980). 

\begin{figure}
\resizebox{8.5cm}{4.5cm}{\includegraphics{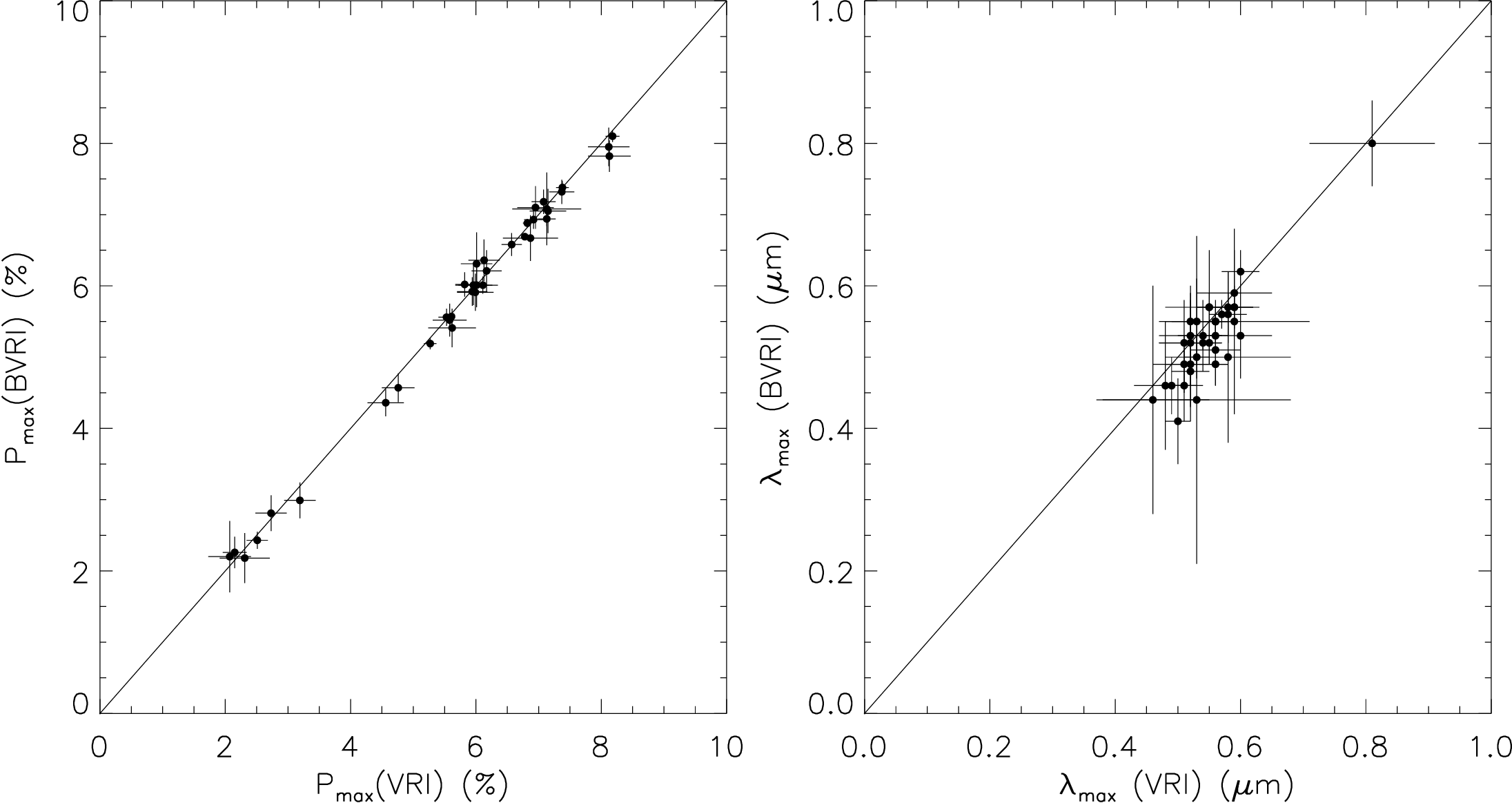}}
\vskip.1cm
\caption{{\it Left panel:} $P_{max}$ computed using four passbands (using K=1.66$\lambda_{max}$) versus the same parameter 
but using three passband data (using K=1.15). {\it Right panel:} Same as left panel but for $\lambda_{max}$. Only 37 stars are used 
which have four ($BV(RI)_{C}$) pass bands data (cf. Table \ref{BVRI_serkdat}).}
\label{3fil4fil}
\end{figure}

\begin{figure*}
\centering
\resizebox{14cm}{15cm}{\includegraphics{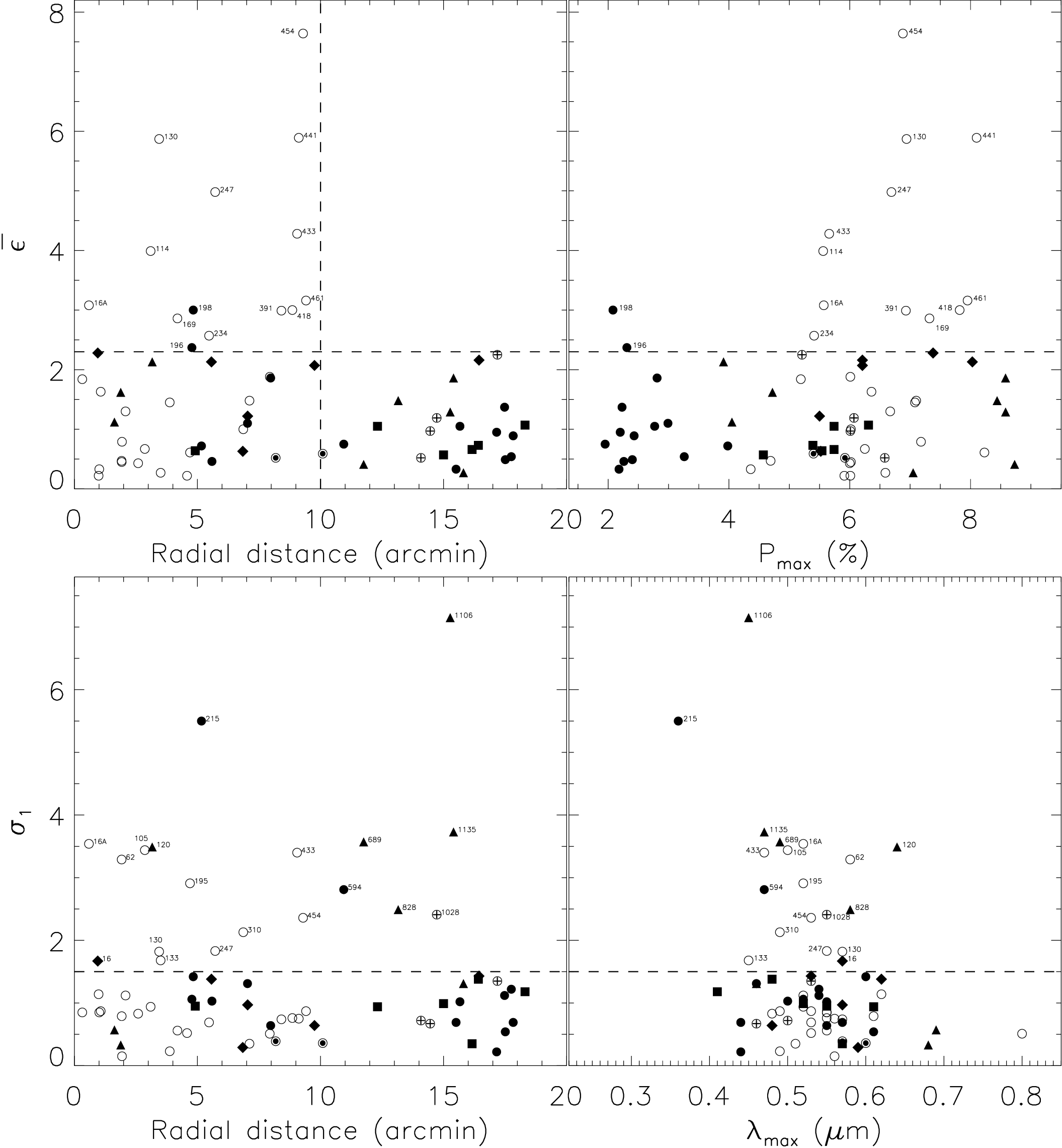}}
\vskip.1cm
\caption{{\it Upper panels}: $\overline{\epsilon}$ versus radial distance of all stars from the center of Be 59 and 
$\overline{\epsilon}$ versus $P_{max}$. {\it Lower panels}: $\sigma_{1}$ versus radial distance of all stars from 
the center of Be 59 and $\sigma_{1}$ versus $\lambda_{max}$. Cluster members are shown with open circles, foreground non-members 
with filled circles, stars in 3rd group with filled squares, late type stars with filled diamonds, 
other non-members with filled triangles, probable members 
with encircled filled circles and stars with uncertainty in their membership with encircled plus symbols.}
\label{raddistri_eps_and_eps_vs_pmax}
\end{figure*}

\begin{table*}
\caption{The $P_{max}$, $\lambda_{max}$, $\sigma_{1}$ and $\overline\epsilon$ for the observed 69 stars}
\label{BVRI_serkdat}
\begin{tabular}{ccccccccccccc} \hline 
 \multicolumn{5}{|c|}{37 stars with BV$(RI)_C$ pass-band data} & \multicolumn{2}{|c|}{} & \multicolumn{5}{|c|}{32 stars with V$(RI)_C$ pass-band data} \\
\hline
Star ID$^\star$ &    $P_{max}\pm\epsilon $ & $\lambda_{max}\pm\epsilon $ &  $\sigma_{1}$ & $\overline\epsilon$ &    &    &  Star ID $^\star$ & $P_{max}\pm\epsilon $ & $\lambda_{max}\pm\epsilon $ &  $\sigma_{1}$ & $\overline\epsilon$ \\
   &    (per cent)           &   $(\mu m)$   &  &  &    &     &    &    (per cent)   &   $(\mu m)$  &    &           \\
(1)     &  (2)   &   (3)   &  (4)  &  (5) & &  & (6)   &   (7)      &    (8)      &      (9)     &     (10)               \\
\hline \hline
     3 &  5.19 $\pm$ 0.08  & 0.55$\pm$  0.02  & 0.85 &  1.84    & & &    47 & 4.05 $\pm$  0.28 & 0.69 $\pm$  0.14 & 0.57 & 1.12    \\  
    15 &  6.01 $\pm$ 0.12  & 0.62$\pm$  0.03  & 1.14 &  0.22    & & &    59 & 4.72 $\pm$  0.31 & 0.68 $\pm$  0.13 & 0.33 & 1.62    \\  
    16 &  7.38 $\pm$ 0.09  & 0.57$\pm$  0.02  & 1.67 &  2.28    & & &    62 & 4.69 $\pm$  0.14 & 0.58 $\pm$  0.03 & 3.29 & 0.47    \\  
   16A &  5.57 $\pm$ 0.04  & 0.52$\pm$  0.01  & 3.54 &  3.08    & & &    64 & 6.02 $\pm$  0.20 & 0.61 $\pm$  0.04 & 0.79 & 0.45    \\  
    20 &  4.36 $\pm$ 0.19  & 0.55$\pm$  0.06  & 0.85 &  0.33    & & &    92 & 6.00 $\pm$  1.10 & 0.48 $\pm$  0.10 & 0.83 & 0.43    \\  
    23 &  6.36 $\pm$ 0.29  & 0.53$\pm$  0.06  & 0.87 &  1.63    & & &   105 & 6.25 $\pm$  0.24 & 0.50 $\pm$  0.02 & 3.44 & 0.67    \\  
    66 &  7.18 $\pm$ 0.17  & 0.56$\pm$  0.03  & 0.15 &  0.79    & & &   120 & 3.91 $\pm$  0.16 & 0.64 $\pm$  0.07 & 3.49 & 2.13    \\  
    69 &  6.67 $\pm$ 0.32  & 0.52$\pm$  0.06  & 1.12 &  1.30    & & &   133 & 6.59 $\pm$  1.03 & 0.45 $\pm$  0.07 & 1.68 & 0.27    \\  
   114 &  5.56 $\pm$ 0.12  & 0.52$\pm$  0.03  & 0.94 &  3.99    & & &   195 & 8.23 $\pm$  0.28 & 0.52 $\pm$  0.02 & 2.91 & 0.61    \\  
   130 &  6.94 $\pm$ 0.10  & 0.57$\pm$  0.02  & 1.82 &  5.87    & & &   196 & 2.31 $\pm$  0.20 & 0.52 $\pm$  0.06 & 1.06 & 2.37    \\  
   157 &  7.08 $\pm$ 0.51  & 0.49$\pm$  0.08  & 0.23 &  1.45    & & &   198 & 2.08 $\pm$  0.14 & 0.60 $\pm$  0.09 & 1.42 & 3.00    \\  
   169 &  7.32 $\pm$ 0.17  & 0.55$\pm$  0.03  & 0.56 &  2.86    & & &   202 & 5.54 $\pm$  0.38 & 0.55 $\pm$  0.05 & 0.95 & 0.64    \\  
   186 &  5.91 $\pm$ 0.26  & 0.53$\pm$  0.05  & 0.52 &  0.22    & & &   215 & 3.98 $\pm$  0.78 & 0.36 $\pm$  0.05 & 5.50 & 0.72    \\  
   234 &  5.41 $\pm$ 0.27  & 0.53$\pm$  0.07  & 0.69 &  2.57    & & &   240 & 8.03 $\pm$  0.25 & 0.62 $\pm$  0.05 & 1.38 & 2.13    \\  
   239 &  2.26 $\pm$ 0.22  & 0.50$\pm$  0.12  & 1.03 &  0.46    & & &   317 & 5.50 $\pm$  0.19 & 0.57 $\pm$  0.03 & 0.97 & 1.22    \\  
   247 &  6.69 $\pm$ 0.04  & 0.55$\pm$  0.01  & 1.83 &  4.98    & & &   433 & 5.66 $\pm$  0.33 & 0.47 $\pm$  0.03 & 3.40 & 4.28    \\  
   307 &  5.52 $\pm$ 0.23  & 0.59$\pm$  0.06  & 0.29 &  0.63    & & &   503 & 5.40 $\pm$  0.20 & 0.60 $\pm$  0.05 & 0.36 & 0.59    \\  
   310 &  6.02 $\pm$ 0.17  & 0.49$\pm$  0.03  & 2.13 &  1.00    & & &   594 & 1.95 $\pm$  0.10 & 0.47 $\pm$  0.03 & 2.81 & 0.75    \\  
   316 &  2.99 $\pm$ 0.25  & 0.46$\pm$  0.09  & 1.31 &  1.10    & & &   689 & 8.73 $\pm$  0.39 & 0.49 $\pm$  0.02 & 3.57 & 0.41    \\  
   324 &  7.10 $\pm$ 0.30  & 0.51$\pm$  0.05  & 0.35 &  1.48    & & &   742 & 5.74 $\pm$  0.29 & 0.61 $\pm$  0.07 & 0.94 & 1.05    \\  
   367 &  6.01 $\pm$ 0.31  & 0.80$\pm$  0.06  & 0.51 &  1.88    & & &   828 & 8.44 $\pm$  0.10 & 0.58 $\pm$  0.01 & 2.49 & 1.48    \\  
   368 &  2.81 $\pm$ 0.25  & 0.55$\pm$  0.13  & 0.64 &  1.86    & & &  1028 & 6.07 $\pm$  0.18 & 0.55 $\pm$  0.02 & 2.41 & 1.19    \\  
   382 &  5.92 $\pm$ 0.20  & 0.57$\pm$  0.05  & 0.39 &  0.52    & & &  1106 & 8.58 $\pm$  0.47 & 0.45 $\pm$  0.02 & 7.15 & 1.29    \\  
   391 &  6.93 $\pm$ 0.13  & 0.57$\pm$  0.03  & 0.74 &  2.99    & & &  1135 & 8.58 $\pm$  0.55 & 0.47 $\pm$  0.03 & 3.73 & 1.86    \\  
   418 &  7.82 $\pm$ 0.22  & 0.55$\pm$  0.04  & 0.76 &  3.00    & & &  1190 & 2.77 $\pm$  0.16 & 0.55 $\pm$  0.05 & 1.02 & 1.05    \\  
   441 &  8.10 $\pm$ 0.08  & 0.56$\pm$  0.02  & 0.75 &  5.89    & & &  1273 & 5.74 $\pm$  0.39 & 0.57 $\pm$  0.08 & 0.35 & 0.66    \\  
   454 &  6.88 $\pm$ 0.06  & 0.53$\pm$  0.01  & 2.36 &  7.64    & & &  1305 & 5.39 $\pm$  0.56 & 0.48 $\pm$  0.06 & 1.38 & 0.73    \\  
   461 &  7.95 $\pm$ 0.27  & 0.49$\pm$  0.04  & 0.87 &  3.16    & & &  1312 & 6.21 $\pm$  0.31 & 0.53 $\pm$  0.04 & 1.43 & 2.16   \\   
   486 &  6.21 $\pm$ 0.29  & 0.48$\pm$  0.05  & 0.64 &  2.07    & & &  1451 & 5.21 $\pm$  0.27 & 0.53 $\pm$  0.04 & 1.35 & 2.25    \\  
   936 &  6.58 $\pm$ 0.16  & 0.50$\pm$  0.03  & 0.72 &  0.52    & & &  1513 & 2.23 $\pm$  0.17 & 0.54 $\pm$  0.07 & 1.12 & 1.37    \\  
   988 &  6.01 $\pm$ 0.28  & 0.46$\pm$  0.05  & 0.67 &  0.97    & & &  1523 & 2.40 $\pm$  0.15 & 0.61 $\pm$  0.09 & 0.54 & 0.49    \\  
  1067 &  4.57 $\pm$ 0.20  & 0.52$\pm$  0.06  & 0.99 &  0.57    & & &    1569 & 3.26 $\pm$  0.20 & 0.54 $\pm$  0.05 & 1.22 & 0.54    \\
  1154 &  2.18 $\pm$ 0.35  & 0.44$\pm$  0.16  & 0.69 &  0.33    & & &          & & & &                                                         \\
  1213 &  7.05 $\pm$ 0.31  & 0.46$\pm$  0.04  & 1.31 &  0.27    & & &          & & & &                                                          \\ 
  1446 &  2.20 $\pm$ 0.50  & 0.44$\pm$  0.23  & 0.22 &  0.95    & & &          & & & &                                                        \\ 
  1580 &  2.43 $\pm$ 0.12  & 0.57$\pm$  0.08  & 0.69 &  0.89    & & &          & & & &                                                    \\
  1684 &  6.31 $\pm$ 0.44  & 0.41$\pm$  0.06  & 1.18 &  1.07    & & &          & & & &                                                    \\
\hline \hline                                                                                                    
\end{tabular}\\
$^\star:$ From P08\\
\end{table*}


\begin{figure}
\begin{small}
\resizebox{8.0cm}{9.0cm}{\includegraphics{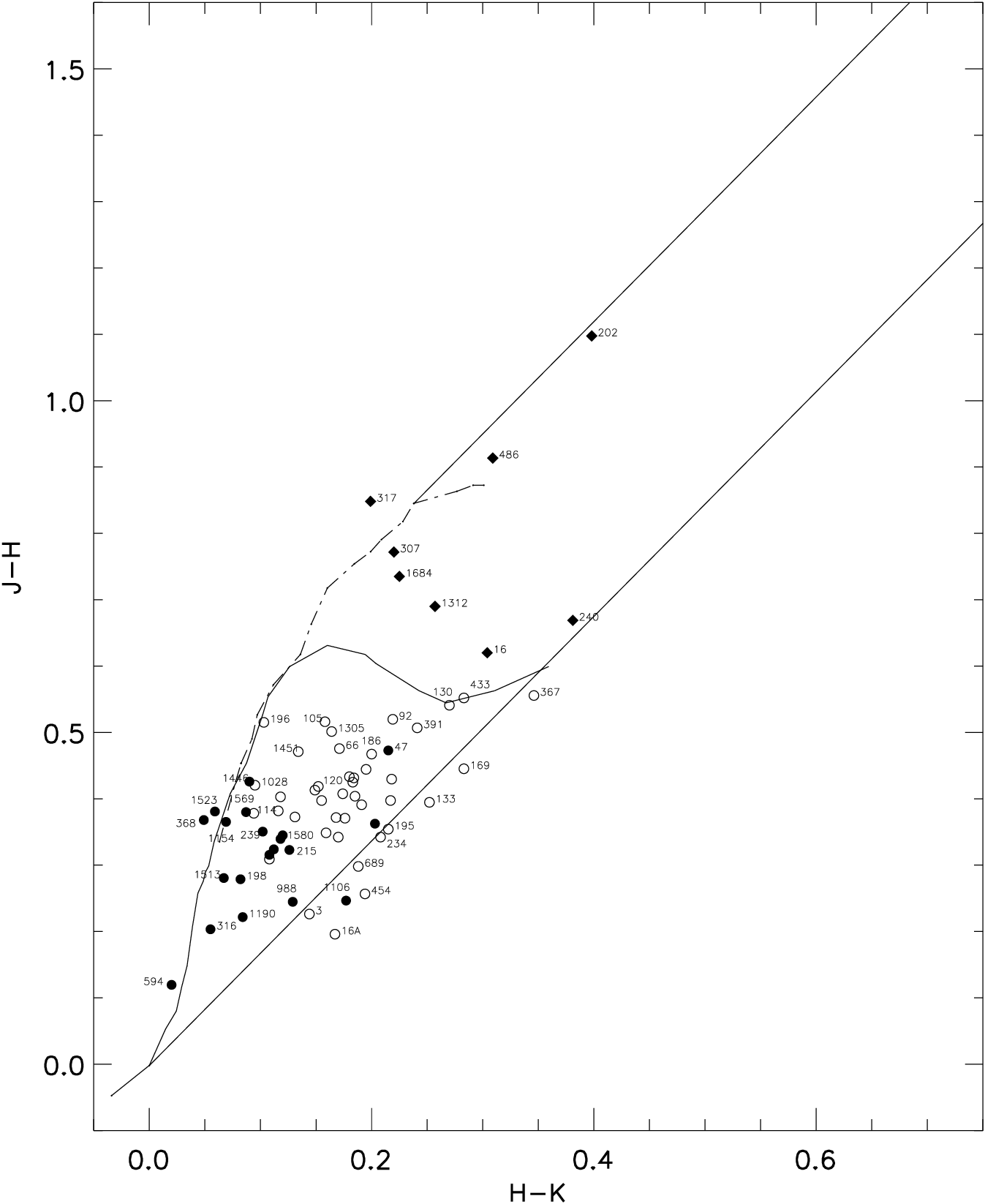}}
\vskip.1cm
\caption{$(J-H)$ versus $(H-K)$ colour-colour diagram for all the observed stars towards Be 59. 
The data are taken from the Two-Micron All-Sky Survey (2MASS)
Point Source Catalog (Cutri et al. 2003). 2MASS data have been converted
into the California Institute of Technology system using the relations provided 
by Carpenter (2001). The theoretical tracks for dwarfs and giants are
drawn (Bessell \& Brett 1988). Reddening vectors are also drawn (Cohen et al. 1981).
The symbols are same as in that of the Fig. \ref{UBBVCCD}.}
\label{NIRCCD}
\end{small}
\end{figure}

\section{DUST PROPERTIES}
\label{dustproperties}

The wavelength dependence of polarization towards many Galactic directions follows the empirical 
relation (Serkowski  et al. 1975; Coyne, Gehrels \& Serkowski 1974; Wilking, Lebofsky, \& Rieke 1982) 
\begin{equation}
 P_{\lambda} = P_{max}~\exp[-K ~$ln$^{2} (\lambda_{max}/\lambda)]
\end{equation}
where $P_{\lambda}$ is the percentage polarization at wavelength $\lambda$ and $P_{max}$ is the 
peak polarization, occurring at wavelength $\lambda_{max}$. The $\lambda_{max}$  is a function of 
the optical properties and characteristic particles size distribution of aligned grains (Serkowski, 
Mathewson \& Ford, 1975; McMillan, 1978). The value of $P_{max}$ is determined by the column 
density, the chemical composition, size, shape, and degree and orientation of the dust grains. The 
parameter $K$, an inverse measure of the width of the polarization curve, was treated as a constant 
by Serkowski et al. (1975), who adopted a value of 1.15 for all the stars. The Serkowski relation with 
$K$=1.15 provides an adequate representation of the observations of interstellar polarization 
between wavelengths 0.36 and 1.0 $\mu$m. 
In one case $P_{max}$ and $\lambda_{max}$  were obtained using the weighted non­linear least square fit to 
the measured polarization by adopting; (1) $K$=1.15 for stars having data in $V(RI)_{C}$ passbands,
or (2) $K$=1.66 $\lambda_{max}$ (Whittet et al. 1992) for stars having data in $B,V(RI)_{C}$ 
passbands. Table \ref{BVRI_serkdat} lists the $P_{max}$, $\lambda_{max}$, $\sigma_{1}$ and 
$\overline\epsilon$ for 69 stars. The estimated values of the $P_{max}$ and $\lambda_{max}$ 
using $B,V(RI)_{C}$ are listed in the second and 
third columns and those with $V(RI)_{C}$ passband data are in 
the seventh and eighth columns. We also computed the 
parameters $\sigma_{1}$\symbolfootnote[2]{The values of $\sigma_{1}$ for each star are computed using the 
expression $\sigma_{1}^{2}=\sum(r_{\lambda}/\epsilon_{p\lambda})^{2}/(m-2)$; where $m$ is the 
number of colours and $r_{\lambda}=P_{\lambda} ~­P_{max} \exp[-K~ln^{2} (\lambda_{max}/\lambda)
$.}, (the unit weight error of the fit for each star) which quantifies the departure of the data from 
standard Serkowski’s law and $\overline{\epsilon}$, 
the dispersion of the polarization angle for each star 
normalized by the average of the polarization angle errors (cf. Marraco, Vega, \& Vrba 1993).  The estimated 
values of $\sigma_{1}$ and $\overline{\epsilon}$ using $BV(RI)_{C}$ 
passband data are listed in the fourth and fifth columns 
whereas those using $V(RI)_{C}$ passband data are in the ninth and tenth columns of Table \ref{BVRI_serkdat}.  
Further, the $P_{max}$ and $\lambda_{max}$ for the 37 stars, which have data in the $BV(RI)_C$ passbands, have been 
calculated using the data of only three passbands $V(RI)_C$ and $K$=1.15. A comparison of the 
$P_{max}$ and $\lambda_{max}$ values obtained using the three and four passband data is shown in Fig. \ref{3fil4fil}, 
which manifests a good agreement. 

If the wavelength dependence of polarization is well represented by the Serkowski law, $\sigma_{1}$ 
should not be greater than 1.5 because of the weighting scheme. 
A higher value ($\textgreater$ 1.5) could be indicative of intrinsic stellar polarization 
(Waldhausen et al. 1999, Feinstein et al. 2008). The polarization angle rotation with wavelength 
($\overline\epsilon$) also indicates the presence of an intrinsic polarization or a change of 
$\lambda_{max}$ along the line of sight (Coyne 1974, Martin 1974).  Systematic variations with 
wavelength in the position angle of the interstellar linear polarization of star light may also be indicative of 
multiple dust layers with different magnetic field orientations along the line of sight (Messinger et al. 
1997). Following the above stated criteria, we consider stars with $\sigma_{1} \textgreater 1.5$ 
and $\overline\epsilon \textgreater 2.3$ as probable candidates to have 
intrinsic polarization and/or polarization angle rotation.  

Fig. \ref{raddistri_eps_and_eps_vs_pmax} plots 
radial distance of stars from the center of the cluster 
versus $\overline\epsilon$ (upper left panel), $\overline\epsilon$ versus $P_{max}$ (upper right panel), 
the radial distance of stars versus $\sigma_1$ (lower left panel) and $\sigma_1$ versus $\lambda_{max}$ 
(lower right panel). 
One can see that a significant number of 
stars show deviation from the normal distribution (14 stars have $\overline\epsilon \textgreater$ 2.3, 
19 stars have $\sigma_1$ $\textgreater$ 1.5 and 5 stars have $\sigma_{1}$ $\textgreater$ 1.5 
as well as $\overline\epsilon \textgreater 2.3$ (cf. Fig. \ref{raddistri_eps_and_eps_vs_pmax} and 
Table \ref{BVRI_serkdat}). It is interesting to mention that the majority of foreground stars 
(12 out of 14) have $\overline\epsilon \textless 2.3$ and all the stars having 
$\overline\epsilon \textgreater 2.3$ are 
located within the cluster region, whereas the majority of the stars (10 stars) 
having $\sigma_1$ $\textgreater$ 1.5 are located within the cluster region. 
About 40 per cent (28 out of 69) stars of the sample show the signatures of either intrinsic polarization or 
rotation in their polarization angles. Ten stars (\#3, 16A, 133, 169, 195, 234, 367, 454, 689 and 1106) 
are found to be located in the near infrared excess zone (see Fig. \ref{NIRCCD}). Six of them (\# 16A, 
133, 195, 454, 689 and 1106) show intrinsic polarization, with $\sigma_{1}$ $\textgreater$ 
1.5. 
The upper right panel of Fig. \ref{raddistri_eps_and_eps_vs_pmax} indicates that  a significant number of stars 
(12 stars) with $P_{max}$ $\textgreater$ 4.5 per cent have $\overline\epsilon$ $\textgreater$ 2.3. 
The star \#247 (BD+66$\degr$ 1673) is located towards the north western edge of the cluster. This star 
was classified previously with objective prism spectra as O9$-$B0  (MacConnell 1968; Walker 1965), 
but is reclassified later as O5 V((f))n (Majaess et al. 2008), making it the hottest star in Cep OB4.  
The star's high temperature drives mass loss via strong stellar 
winds (Yang and Fukui 1992; Gahm et al. 2006). Our polarimetric results indicate an intrinsic 
nature of polarization as it has 
$\sigma_{1}$=1.83 and $\overline{\epsilon}$=4.98. 

Another criterion to detect intrinsic stellar polarization is based on $\lambda_{max}$. A star 
having $\lambda_{max}$ much lower than the average value of the ISM 
(0.545 $\mu$m; Serkowski, Mathewson \& Ford 1975) is considered as a candidate to have an intrinsic component of 
polarization (Orsatti, Vega, \& Marraco 1998). In the present study only one star 
\# 215 has been found to have a much lower value of $\lambda_{max}$=0.36$\pm$0.05$\mu$m. This star has $\sigma_{1}$=5.5. 

\subsection{Extinction law}

\begin{figure*}
\begin{small}
\resizebox{16cm}{13cm}{\includegraphics{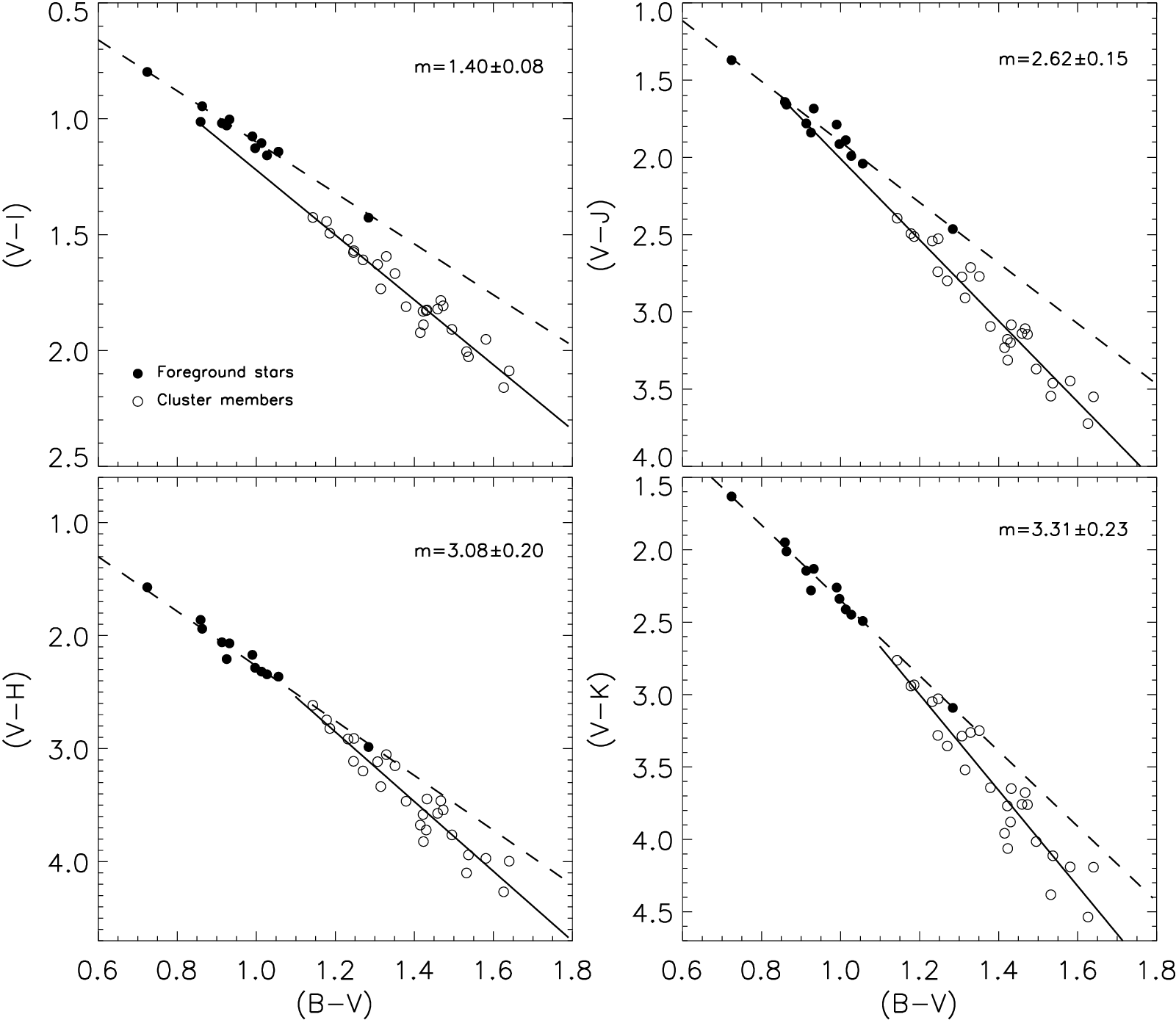}}
\vskip.1cm
\caption{$(V-I)$, $(V-J)$, $(V-H)$, $(V-K)$ versus $(B-V)$ two-colour diagrams for the observed stars towards Be 59.
The first group of foreground stars are shown with filled circles, which follow the normal reddening by 
the general diffuse ISM (see Table 3 of Pandey et al. 2003). The cluster members (open circles) 
show anomalous reddening as their slopes are significantly different from those of foreground stars. 
The fitted slopes for cluster members at each wavelength combination are over laid on the figure.}
\label{DM_EBV_and_VI_BV}
\end{small}
\end{figure*}

To study the nature of the extinction law in the cluster region we used the TCDs as described 
by Pandey et al. (2000, 2003), in the form of ($V-\lambda$) vs. ($B-V$), 
where $\lambda$ is one of the wavelengths of the broadband filters $R,I,J,H,K$ or $L$, 
to separate the influence of the normal extinction produced by the diffuse 
ISM from that of the abnormal extinction arising within regions having a peculiar 
distribution of dust sizes (cf. Chini \& Wargau 1990, Pandey et al. 2000). The ($V-\lambda$) vs. 
($B-V$) TCDs for the cluster region are shown in Fig. \ref{DM_EBV_and_VI_BV}. 
The distribution of field stars follows a normal reddening law 
for the foreground inter-stellar matter.  The slopes of the distributions for cluster members, 
$m_{cluster}$, are found to be $1.40\pm0.08$, $2.62\pm0.15$, $3.08\pm0.20$, 
and $3.31\pm0.23$ for $(V-I)$, $(V-J)$, $(V-H)$, $(V-K)$  versus $(B-V)$ TCDs respectively. 
The ratios ${E(V-\lambda)}\over {E(B-V)}$ and the ratio of total-to-selective extinction in the cluster 
region, $R_{cluster}$, is derived using the procedure given by Pandey et al. (2003). Assuming the 
value of $R_V$ for the diffuse foreground ISM as 3.1, the ratios ${E(V-\lambda)}\over {E(B-V)}$ yield 
$R_{cluster} = 4.0\pm0.1$, which indicates an anomalous reddening law.  P08 have 
also estimated an anomalous reddening law in the cluster region with $R_{cluster} = 3.7 \pm 0.3$, with a 
normal reddening law for the foreground diffuse ISM. In the central region of Be 59, MacConnell (1968) 
also found evidence for a large value (3.4$-$3.7) of $R_V$.
Several studies have already pointed out an anomalous reddening law with a high $R_V$ value in the vicinity of 
star-forming regions (see e.g. Pandey et al. 2003 and references therein), however for the Galactic 
diffuse ISM a normal value of $R_V=3.1$ is well accepted. The higher than the normal 
values of $R_V$ has been attributed to the presence of larger dust grains. There is evidence that within 
dark clouds accretion of ice mantles on grains and coagulation due to colliding grains change the 
size distribution towards larger particles. On the other hand, in star-forming regions, radiation from 
massive stars may evaporate ice mantles resulting in small particles. Here it is interesting to mention 
that Okada et al. (2003), on the basis of the [\Siii] 35 $\mu$m to [\Nii] 122 $\mu$m ratio, suggested 
that efficient dust destruction is occurring in the ionized region of Be 59. Chini \& Kruegel (1983) and 
Chini \& Wargau (1990) have shown that either larger or smaller grains  may increase the ratio of 
total-to-selective extinction.

The weighted mean value of $\lambda_{max}$ for the cluster region was estimated to be 0.538$\pm$0.004$\mu$m. 
The mean value, within error, is comparable to the value measured in the general ISM 
(0.545 $\mu$m, Serkowski et al. 1975). Using the relation $R_V = (5.6 \pm 0.3) \lambda_{max}$ 
(Whittet \& Van Breda 1978), the value of $R_V$, the total-to-selective extinction, comes out to be 
3.01$\pm$0.16, which is in agreement with the average value ($R_V$ = 3.1) for the Milky Way, 
but is in contradiction with the result 
obtained from $(V-\lambda)/ (B-V)$ TCDs. The mean value of $\lambda_{max}$ for 
foreground stars is estimated as 0.498$\pm$0.017$\mu$m, which 
yields the $R_V$ value as 2.79$\pm$0.18 for the foreground diffuse ISM. This indicates a smaller 
value of $R_V$ for the diffuse ISM towards the direction of the Be 59 ($l=118.2\degr$). 
There is much evidence in the literature that indicates significant variations in the properties of 
interstellar extinction along various Galactic directions. Whittet (1977) reported that the value of 
$R_V$ in the Galactic plane can be represented by a sinusoidal function of the form 
$R_V$ = 3.08+0.17 $\sin$($l$ + 175$\degr$), which indicates a minimum value of $R_V$ at 
$l \sim 95\degr$. The above relation suggests a value of $R_V \sim 2.9$ towards
the direction of Be 59 (i.e. $l=118.2\degr$). The study by Geminale \& Popowski (2004)
also suggests a lower value ($\sim 2.9$) towards Galactic longitude $l \sim 120\degr$.
Thus the present estimation of $\lambda_{max}$, and consequently the value of $R_V$ for the diffuse ISM is 
in agreement with the values reported in the literature. Hence we conclude that the $R_V$ 
for the intra-cluster matter could be higher in comparison to that for the general diffuse matter 
towards the direction of Be 59. Here it is worthwhile to note that there is much evidence 
of variation of grain size-distribution towards the direction of Be 59.

\subsection{POLARIZATION EFFICIENCY}

\begin{figure*}
\resizebox{15cm}{15cm}{\includegraphics{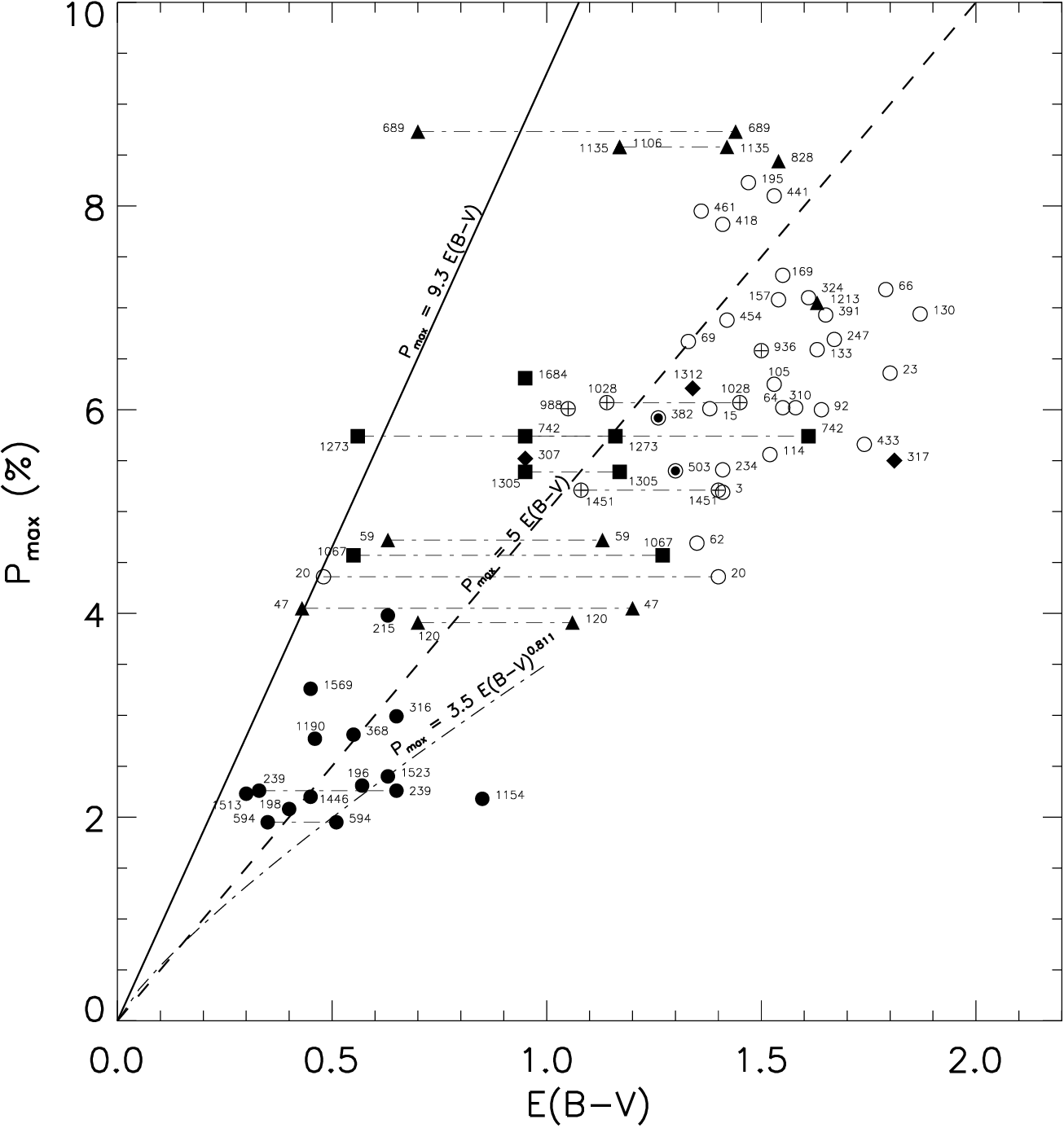}}
\vskip.1cm
\caption{$P_{max}$ versus $E(B-V)$ for the stars with available reddening (cf. Table \ref{BVRI_serkdat}).
The symbols are same as that of the Fig. \ref{raddistri_eps_and_eps_vs_pmax}. The stars with two possible reddening values are
connected by thin dash-dotted line.The solid line represents the empirical upper limit relation for
the polarization efficiency (by assuming $R_V$=3.1) of $P_{max}$ = 9.3 $\times$ $E(B-V)$ (Serkowski et al. 1975).
The dashed line represents the relation $P_{max}$ = 5 $\times$ $E(B-V)$ (Serkowski et al. 1975)
and the dash-dotted line represents the relation $P_{max}$ = 3.5 $\times$ $E(B-V)^{0.8}$ by Fosalba et al. (2002).}
\label{poleff}
\end{figure*}

The ratio of $P_{max}/E(B-V)$ is known to be a measure of the polarization efficiency of the ISM and it 
depends mainly on the grain alignment efficiency, the magnetic field strength and the amount of 
depolarization due to the radiation traversing clouds with different magnetic field 
directions. When the light passes through multiple dust layers, the resultant polarization may 
increase or decrease depending upon the orientation of the magnetic field in each dust layer 
(see for example, Feinstein et al. 2003b, Orsatti et al. 2003, Mart\'{i}nez et al. 2004, 
Vergne et al. 2007, 2010 and E11)
The observed polarization and extinction data towards a particular direction of the Galaxy 
provides important inputs to test the models dealing  with the extinction and alignment of the grains. 
Figure \ref{poleff} displays the polarization efficiency diagram for the observed stars. 
The symbols are the same as in Fig. \ref{raddistri_eps_and_eps_vs_pmax}. It is well known that for 
the diffuse ISM the polarization efficiency can not exceed the empirical upper limit given by, 
$P_{max}$=3 $A_V$ $\simeq$ 3 $R_V$ $E(B-V)$ $\simeq$ 9.3 $E(B-V)$ per cent 
(assuming $R_V$=3.1, Hiltner \& Johnson 1956; Serkowski et al. 1975) 
and the same is shown  by a continuous line in Fig. \ref{poleff}. For the average ISM, 
Serkowski et al. (1975) have found that the polarization efficiency of the ISM 
follows the mean relation $P_{max}$ $\simeq$ 5 $E(B-V)$, which is shown by a dashed line. 
The recent estimate of the average  polarization efficiency for the general diffuse ISM by 
Fosalba et al. (2002), which is  valid for $E(B-V)$ $\textless$ 1.0 mag, is shown with 
a dash-dotted line. 

Figure \ref{poleff} shows that the foreground stars (filled circles) are distributed 
along the dashed line, which suggests that the dust grains in the dust layer 
located at $\la$ 470 pc have polarization efficiency comparable to the 
average polarization efficiency  ($\sim$5 per cent per mag) of the diffuse ISM. 
The stars located at $\sim$ 700 pc are shown with filled squares. The colours of these 
stars indicate two values of $E(B-V)$  (see Table \ref{ebv_memb_results}). We have used 
both values of $E(B-V)$ for these stars and those data are connected with thin dashed-dotted lines. 
In general it seems that the dust layer located at 500 $\la$ d $\la$ 800 pc also has an average 
polarization efficiency (i.e., 5 per cent per mag). The majority of the cluster members (open circles) are 
distributed below the dashed line thereby indicating that the intra-cluster medium exhibits 
less polarization efficiency than the mean value for the diffuse ISM ($\sim$ 5 per cent per mag). 
The large dispersion in $P_{max}$ (4$-$8 per cent) for cluster members is compatible with
the differential reddening within the cluster ($\sim$ 1.4$-$1.7 mag).

The net polarization due to the intra-cluster medium is estimated to be $\sim$2.2 per cent 
(cf. Table \ref{dust_layer_properties}). The differential $E(B-V)$ due to the intra-cluster medium 
is $\sim$0.3 mag. Thus the net polarization efficiency due to the intra-cluster medium comes out to be 
$\sim$ 7.3 per cent per mag, which is higher than that due to the diffuse ISM. The 
small dispersion in the mean value of $\theta_V$ (4$\degr$, Table \ref{mean_pv_tv_m_nm_2ndgrp}) of 
cluster members also indicates a better alignment of dust grains in the intra-cluster medium. 
As discussed in Sec \ref{dust_distri}, for the foreground stars of the first 
group and third group as well as the cluster members, the mean polarization angle 
changes with increasing distance systematically 
which may lead to the depolarization effect in the case of radiation from cluster members. 
Hence the less polarization efficiency of the 
intra-cluster medium, as seen in Fig. \ref{poleff}, could be because 
of different alignment of dust grains in the foreground dust layers.
Similar kind of depolarization effect has been observed towards 
Trumpler 27 by Feinstein et al. (2000), Hogg 22 and NGC 6204 by Mart\'{i}nez, Vergne and Feinstein (2004) and 
towards NGC 6124 by Vergne et al. (2010). Here it is interesting to mention that stars \# 418, 441 and 461, located 
at the northern periphery of the cluster have the highest polarization efficiency among the cluster 
members. These stars have polarization efficiencies greater than the average value (5 per cent per mag) for 
the diffuse ISM. Four stars \# 689, 828, 1106 and 1135 are located outside the boundary of the cluster 
and towards north-east of the cluster center. These stars have the highest polarization ($P_V$ $\sim$ 7.6$-$8.7 per cent) 
and polarization efficiency greater than 5 per cent per mag. The $E(B-V)$ values of these stars range $\sim$1.20$-$1.50 mag. 
Interestingly, all these four stars have their $\sigma_{1}$ $\textgreater$ 1.5 
(cf. Fig. \ref{raddistri_eps_and_eps_vs_pmax}), thereby indicating presence of 
intrinsic polarization.

\begin{figure*}
\resizebox{13cm}{22cm}{\includegraphics{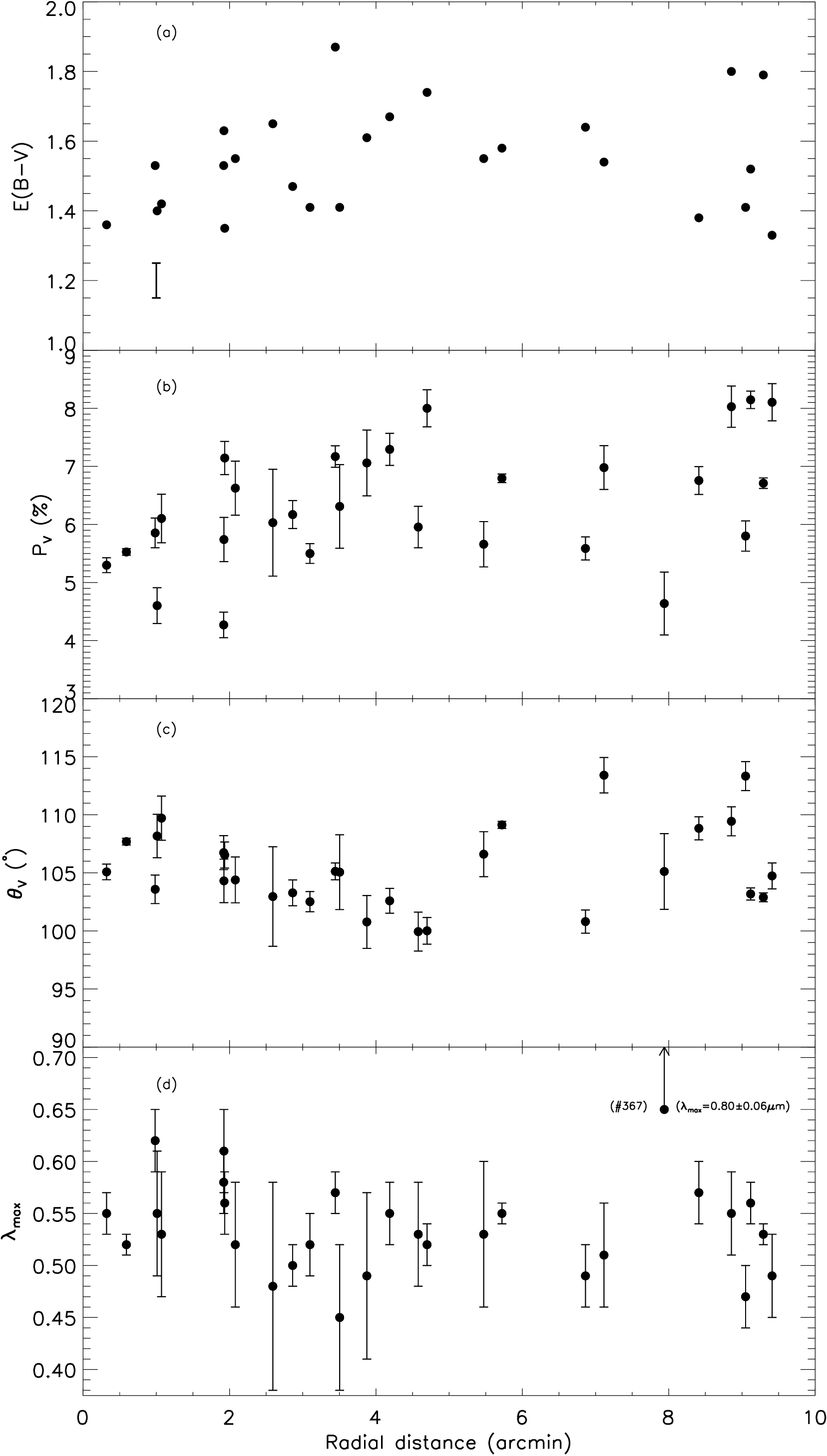}}
\vskip.1cm
\caption{Radial variation of $E(B-V)$, $P_V$, $\theta_V$ and $\lambda_{max}$ for only the cluster members. In the
top panel, only the cluster members with $E(B-V)$ are used. The estimated error (0.05 mag) in $E(B-V)$ is also
plotted for a reference.}
\label{radvar_ebv_pv_tv_lmax}
\end{figure*}

\section{Spatial variation of $E(B-V)$, $P_V$, $\theta_V$ and $\lambda_{max}$}

Fig. \ref{radvar_ebv_pv_tv_lmax}(a) shows the spatial distribution of $E(B-V)$ as a function of 
radial distance from the center of the cluster. The distribution reveals a lack of reddening material 
near the center of the cluster. The $E(B-V)$ increases away from the center. 
The $E(B-V)$ values reaches its maximum at $\sim$ 4$\arcmin$ ($\sim$ 1.2 pc) and remains constant 
up to $\sim$ 6$\arcmin$ (1.8 pc). The $E(B-V)$ value decreases for radial distances $\textgreater$ 6$\arcmin$. 
The spatial distribution suggests that the density of the reddening material is high at $\sim$ 1.2$-$1.6 pc in 
comparison to that near the cluster center. On the basis of one-dimensional raster-scan 
observations of Sh171 (Be 59) region, Okada et al (2003) have studied the the spatial distribution of 
line intensities, representing the lowly-ionized gas, the highly ionized gas and photo dissociation region 
(gas phases) as a function of radial distance from the cluster. The lines representing 
the neutral region (e.g., [\Oi] 63$\mu$m, 146$\mu$m; [\Cii] 158$\mu$m; [\Siii] 35$\mu$m), 
lowly ionized region ([\Nii] 122$\mu$m) and highly ionized region ([\Oiii] 52$\mu$m, 88$\mu$m; [\Niii] 57$\mu$m) 
show maxima at $\sim$ 1 pc and $\sim$ 4 pc indicating the presence of high density gas 
at these positions. 
Their figure 6, showing 0.61 GHz continuum flux on the line along the raster-scan observations, 
also indicated that the flux increases systematically with radial distance from the cluster 
center and reaches the maximum at $\sim$ 1.2 pc and remains constant up to $\sim$ 1.7 pc. 
The flux shows a decreasing trend for radial distance $\ga$ 1.7 pc. The spatial distribution of 
0.61 GHz continuum flux agrees nicely with the $E(B-V)$ distribution as shown in Fig. \ref{radvar_ebv_pv_tv_lmax}(a). 
Thus, it is clear that the density of the gas increases with the radial 
distance from the cluster center with the maximum at $\sim$ 1 pc. 

The spatial variation of $P_V$ as a function of 
radial distance from the cluster center (Fig. \ref{radvar_ebv_pv_tv_lmax}(b)) shows a systematic increasing 
trend with the radial distance up to $\sim$ 5$\arcmin$; however for radial distance 
$\ga$5$\arcmin$, the distribution of $P_V$ shows no trend but a significant scatter. 
The variation of $P_V$ agrees well with the distribution of gas as discussed above, indicating 
that the average polarization efficiency in the core as well as in the corona of the cluster is the same. 
Fig. \ref{radvar_ebv_pv_tv_lmax}(c) shows the distribution of $\theta_V$ as a function of 
radial distance. The $\theta_V$ distribution also shows a systematic change with the radial 
distance, in the same sense that the average $\theta_V$ value ($\sim$ 110$\degr$) at the center 
decreases to $\sim$100$\degr$ at $\sim$5$\arcmin$. For radial distances $\ga$5$\arcmin$, 
the distribution of $\theta_V$ shows a scattered distribution around 102$\degr$. On the basis of 
the $\theta_V$ distribution we can conclude that the magnetic field orientation in the core of the cluster 
is significantly different from that in the coronal region, which is comparable to the magnetic field 
orientation of the third group of the stars ($\sim$ 96$\degr$). We presume this 
could be either due to the molecular cloud at the center being perturbed during the cloud collapse 
or to strong stellar winds/supernova explosion. 

Fig. \ref{radvar_ebv_pv_tv_lmax}(d) shows the spatial variation of $\lambda_{max}$ obtained 
for identified members of the cluster, which manifests that the $\lambda_{max}$ is higher near 
the center of the cluster. The average value of $\lambda_{max}$ 
decreases up to $\sim$ 5$\arcmin$ with increasing radial distance from the cluster center. 
The distribution of $\lambda_{max}$ suggests that the dust grain size near the center of the cluster 
is relatively higher in comparison to that in the coronal region. Okada et al (2003), on the basis of 
the [\Siii] 35$\mu$m to [\Nii] 122$\mu$m ratio, suggested that the efficient dust destruction is 
occurring in the ionized region. It is possible that the smaller dust grains might have been evacuated 
from the central region of the cluster due to strong stellar winds leaving relatively large  dust grains 
in the central region of the cluster. 
\section{Conclusions}
In the present study we have carried out polarimetric observations towards the 
direction of young open cluster Be 59 in the $B, V, (R, I)_C$ bands. The aim of the study 
was to investigate the properties of dust grains in the ISM towards the direction of Be 59 as well as the 
properties of intra-cluster dust. \\

Following are the main conclusions of the present study:\\

The distribution of $P_V$ and $\theta_V$ suggests three dust layers towards the 
direction of Be 59 at $\sim$300 pc, $\sim$500 pc and $\sim$700 pc respectively. The total polarization 
due to these dust layers is found to be $\sim$0.2$-$1.0, $\sim$1.0$-$3.0 and $\sim$ 5.5 per cent respectively. 
The magnetic field orientation in these dust layers is different from each other. The magnetic field 
orientation of the first dust layer ($\sim$ 82$\degr$) is rather similar to that of the GP (86$\degr$). 

We have further shown that the polarization measurements in combination with the $(U-B)-(B-V)$ colour-colour 
diagram provide a better estimation of the cluster members. The polarization measurements of the identified 
cluster members reveal that the net polarization due to the intra-cluster medium is estimated to be $\sim$2.2 per cent.
About 40 per cent cluster members show the signatures of either intrinsic polarization or rotation in their 
polarization angles. 

The TCDs of the identified cluster members reveal an anomalous reddening law in the cluster region. 
The weighted mean values of $\lambda_{max}$ for the cluster region and for the foreground stars are estimated 
as 0.54$\pm$0.01$\mu$m and 0.50$\pm$0.02$\mu$m. The above estimate of $\lambda_{max}$ for the foreground ISM 
indicates relatively smaller dust grain sizes, consequently smaller value of total-to-selective 
absorption $R_V$=2.79$\pm$0.18 in comparison to the normal value for the diffuse ISM ($R_V$=3.1). Thus mean 
$\lambda_{max}$ value for the cluster region (0.54$\pm$0.01$\mu$m) suggests a relatively larger grain size in the cluster 
region in comparison to those in the general diffuse ISM towards the cluster region. 

The foreground dust layers have polarization efficiency comparable to the average polarization efficiency 
($\sim$ 5 per cent per mag) of the diffuse ISM, whereas the majority of the cluster members indicate a smaller polarization 
efficiency for the intra-cluster medium. 
It indicates that the star light of the cluster
members might have been depolarized because of the non-uniform alignment of dust grains in the foreground dust layers
and in the intra-cluster medium. 

The spatial distribution of $E(B-V)$ in the cluster region shows an increasing trend with the radial distance. 
The polarization is also found to be systematically increasing with the radial distance from the cluster center. 
Both $E(B-V)$ and $P_V$ values reach a maximum value at $\sim$4$\arcmin$-5$\arcmin$ ($\sim$ 1.2$-$1.5 pc). 
The $\theta_V$ as well as $\lambda_{max}$ for the cluster members are found to decrease systematically with 
increasing radial distance. 

\section*{Acknowledgments}
The authors are thankful to the referee Prof. Michael S. Bessell for his critical 
reading of the manuscript and several useful suggestions, which greatly improved the scientific 
content of the paper. This publication makes use of data from the 2MASS (a joint project of the
University of Massachusetts and the Infrared Processing and Analysis Center
/California Institute of Technology, funded by the National Aeronautics
and Space Administration and the National Science Foundation).
This research has made use of the WEBDA database, operated  at the
Institute for Astronomy of the University of Vienna, as well as has used
the images from the Digital Sky Survey  (DSS), which
was produced  at the  Space Telescope Science  Institute under  the US
Government grant  NAG W-2166. We have also used  NASA's Astrophysics Data System
and {\small IRAF}, distributed   by  National   Optical  Astronomy
Observatories,  USA. This work is partially supported by Indo-Taiwan S\&T programme.
{}

\end{document}